\documentclass[12pt]{article}
\usepackage{a4wide}
\usepackage{graphicx}
\newcommand{\be}{\begin{equation}}
\newcommand{\ee}{\end{equation}}
\newcommand{\ba}{\begin{eqnarray}}
\newcommand{\ea}{\end{eqnarray}}
\begin{document}
\begin{flushright}
LU-TP 02-24\\
TTP02-08
\end{flushright}
\vspace{0.5cm}

\thispagestyle{empty}
\begin{center}
{\Large\bf Exploring Light-Cone Sum Rules 
for Pion and Kaon Form Factors }
\vskip 1.5true cm

{\large\bf
Johan~Bijnens$\,^{a}$ and Alexander~Khodjamirian$\,^{b\,*)}$}~
\vskip 1true cm

{\it$^a$ Department of Theoretical Physics, Lund University,\\
S\"olvegatan 14A, S - 223 62 Lund, Sweden}\\
\vspace{0.3cm}
{\it$^b$ Institut f\"ur Theoretische Teilchenphysik, Universit\"at
Karlsruhe,\\  D-76128 Karlsruhe, Germany } \\

\end{center}

\vskip 1.0true cm
\begin{abstract}
\noindent

We analyze the higher-twist effects and the $SU(3)$-flavour symmetry breaking 
in the correlation functions used to calculate form factors of
pseudoscalar mesons in the QCD light-cone sum rule approach.
It is shown that the Ward identities for these correlation functions yield
relations between twist-4 two- and three-particle distribution
amplitudes. In addition to the
relations already obtained from the QCD equations of motions,
we have found a new one. With the help of these relations, the 
twist-4 contribution to the light-cone sum rule 
for the pion electromagnetic form factor is reduced to a very simple 
form. Simultaneously, we correct a sign error in the earlier calculation.
The updated light-cone sum rule prediction for the pion 
form factor at intermediate 
momentum transfers is compared with the recent Jefferson Lab data. 
Furthermore, from the correlation functions with strange-quark currents 
the kaon electromagnetic form factor and  the $K\to \pi$ weak transition
form factors are predicted with  $O(m_s)\sim O(m_K^2)$
accuracy.

\end{abstract}

\vspace{1cm}
\noindent $^{*)}${\small \it on leave from Yerevan Physics 
Institute, 375036 Yerevan, Armenia} 

\newpage

\section{Introduction}

An accurate knowledge of the pion and kaon light-cone distribution amplitudes
(DA) introduced in the studies of hard exclusive processes in QCD 
\cite{exclusive}
is important for various frameworks
where these DA are being used. Among the most topical
applications one could mention the calculations 
of exclusive semileptonic
and hadronic $B$-meson transitions into pions and kaons using pQCD \cite{pQCD},
QCD factorization \cite{BBNS} or light-cone sum rules \cite{LCSRB}.
Although there are definite
indications that at the normalization scale of $O(1 \mbox{GeV})$ 
the leading twist 2 pion DA is already quite close 
to its  asymptotic shape, one still encounters a large uncertainty of the
nonasymptotic part. Moreover, very little is known about nonasymptotic 
$SU(3)$-flavour asymmetry in the twist 2 kaon DA.

One of the promising ways to study DA 
is to employ  vacuum-to-pion or vacuum-to-kaon 
correlation functions  of light-quark currents.
At high virtualities, using the operator-product expansion (OPE) 
near the light-cone, these correlation functions are expressed in terms
of DA. On the other hand, the same correlation functions are  
related, via dispersion relations, 
to the observable form factors of pions and kaons 
with the contributions of excited hadronic states 
approximated by  quark-hadron duality. In the resulting relations, 
known as {\em light-cone sum rules} (LCSR) \cite{LCSR}, the 
experimental data on form factors can be used 
to yield nontrivial constraints on DA.
The LCSR for the pion electromagnetic (e.m.) form factor
was derived in Refs.~\cite{BH,BKM} and for the $\gamma^*\gamma\pi^0$ 
form factor in Ref.~\cite{AKpi}. In order to further 
increase the accuracy of these sum rules one has to gain a better
control over higher-twist effects in the OPE. 
In the case of the pion form factors
the most important subleading contribution to the LCSR is of twist 4. 
The kaon e.m. form factor which so far was not analyzed in the LCSR
framework, demands also inclusion of the twist 3 effects proportional 
to $m_s$.   

The aim of this paper is twofold. First, we analyse the higher-twist effects in the vacuum-to-pion and vacuum-to-kaon correlation functions.
We demonstrate that a new useful tool is provided by standard 
Ward identities for the conserved e.m. and axial (in the chiral limit) 
currents. Simultaneously, we correct a sign error in the previous 
calculation of the twist 4 term and update the LCSR prediction for the 
pion e.m. form factor. Second, we include $SU(3)$-flavour symmetry 
breaking effects at $O(m_s)\sim O(m_K^2)$  in the correlation functions. 
We calculate the twist 3 part and 
obtain LCSR for the kaon e.m. and $K\to \pi$  weak transition 
form factors at intermediate spacelike momentum transfers.

The plan of the paper is as follows. In Sect.~\ref{Sect2} we introduce
a generic correlation function, which yields LCSR for the pion,
kaon, and $K\to \pi$ form factors for different flavour combinations
of light-quark currents.
The correlation function
is then calculated with twist 4 accuracy including first-order
in quark mass terms. In Sect.~\ref{Sect3} we derive the Ward identities
in the chiral limit and demonstrate that they lead
to relations between two- and three-particle DA of twist 4.
In Sect.~\ref{Sect4} the
numerical results for the pion form factor are presented
with a corrected twist 4 contribution. A comparison 
of our prediction is made with the recent data
on the pion e.m. form factor obtained at CEBAF.
Sect.~\ref{Sect5}  contains LCSR
results for the kaon electromagnetic form factor, and Sect. \ref{Sect6}
deals with the $K\to \pi$ weak transition form factor.
We summarize our conclusions in Sect.~\ref{Conclusions}.
The appendices contain the expansion of the quark propagator in
App.~\ref{AppA}, the definitions of the DA and their asymptotic expansions
in App.~\ref{AppB} and in App.~\ref{AppC} the $\alpha_S$ corrections
to the twist 2 LCSR obtained in Ref.~\cite{BKM}.

\section{Correlation functions}
\label{Sect2}

As a starting point, we introduce a generic correlation function: 
\be
T_{\mu\nu}(p,q)\! = i\!\int\! d^4 x \,e^{iqx}
\langle 0| T\{\left(\bar{q}_2(0)\gamma_\mu\gamma_5 q_1(0)\right)
\left( e_1\bar{q}_1(x)\gamma_\nu q_1'(x)\!+e_2\bar{q}_2'(x)\gamma_\nu q_2(x)
)\right\}\! |P(p)\rangle\,,
\label{2:cor}
\ee
where, in order to obtain the LCSR for the pion e.m. form factor  
the following quark-flavour combination has to be taken: 
$q_1=q_1'=u$, $q_2=q_2'=d$. In this case the on-shell hadronic state  
$P=\pi^+$, and $e_1=e_u=2/3$, $e_2=e_d=-1/3$ are the quark e.m. 
charges in the units of $e$. To calculate the kaon e.m. form factor, 
one simply has to replace $d\to s$ and $\pi^+\to K^+$ in the above. 
There are two other, 
physically interesting correlation functions yielding two independent 
LCSR for the $K\to \pi$ weak transition form factors obtained 
from  Eq.~(\ref{2:cor}) at $q_1=s$, $q_1'=u$, $q_2=d$, $P=\pi^+$,
$e_1=1$, $e_2=0$ and at $q_1=d$, $q_2=u$, $q_2'=s$, $e_1=0$, $e_2=1$, 
$P=K^0$. Summarizing, if one calculates the correlation
function (\ref{2:cor}) the result can easily be adjusted to any of
the flavour combinations listed above. 
\begin{figure}
\begin{center}
\includegraphics[width=0.9\textwidth]{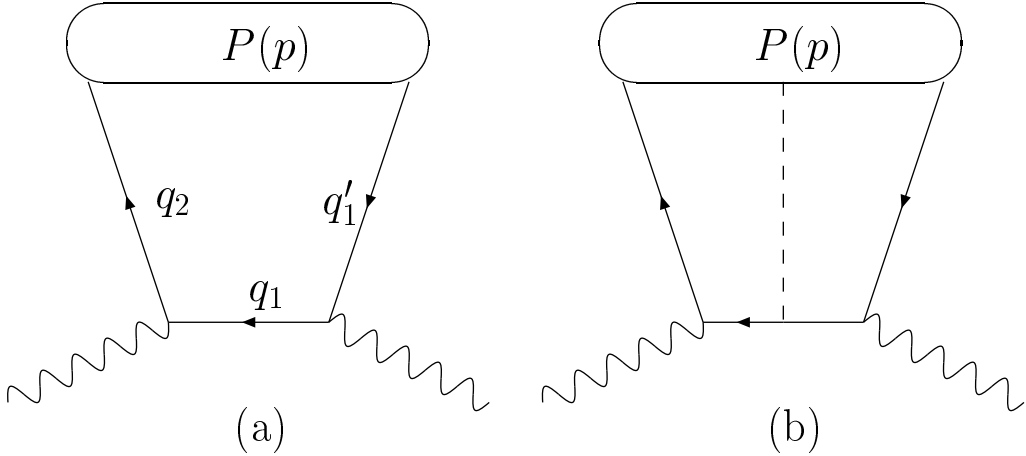}
\end{center}
\caption{
{\it Diagrams corresponding (a) to the
leading-order of the correlation function (\ref{2:cor});
(b) to the contributions of twist 3,4 quark-antiquark-gluon
DA. Solid, dashed, wavy lines and ovals represent quarks, gluons,  
external currents and pseudoscalar meson DA, respectively.}
\label{fig1}}
\end{figure}
If the external 4-momenta squared $q^2$ and $(p-q)^2$ are spacelike
and large, the operator product in the 
correlation function (\ref{2:cor}) can be expanded near the light-cone
in terms of pion or kaon DA of increasing twists. One may then 
retain a few first 
terms in this expansion, having in mind that higher twists are suppressed
by inverse powers of $Q^2=-q^2$ and/or $|(p-q)^2|$ (for a more detailed
discussion see e.g. Refs.~\cite{BKM,CK}).   
There are two leading-order diagrams obtained from the two terms 
in Eq.~(\ref{2:cor}) by contracting the quark fields 
$q_1$ with $\bar{q}_1$ and $q_2$ with $\bar{q}_2$, respectively
and replacing them by the free-quark propagators. 
The first diagram proportional to $e_1$ is depicted in Fig.~\ref{fig1}a. 
The second diagram, proportional to $e_2$  is obtained from the first 
one by changing the direction of the quark line and replacing  
the quark-flavour indices $1\leftrightarrow 2$. 
The next-to-leading approximation for the
quark propagator generates 
the diagram in Fig.~\ref{fig1}b (and its $\sim e_2$ counterpart) 
which brings three-particle quark-antiquark-gluon
DA of twist 3 and 4 into the game. This diagram is calculated
using the first-order in gluon field term in the light-cone 
expansion of the quark propagator given in App.~\ref{AppA}.    
We systematically retain all terms of  $O(m_q)\sim O(m_P^2)$ 
in order to be able to account for $SU(3)$ breaking effects 
in the LCSR for the kaon form factors. At the same time, the $O(m_q^2)$ 
contributions arising, e.g. from the denominators of quark propagators are 
neglected.

The result for the correlation function (\ref{2:cor}) 
obtained to twist 4 accuracy reads
\ba
T_{\mu\nu}(p,q) = if_P\int\limits_0^1 du \Big\{
T_1(Q^2,s,u) p_\mu p_\nu + 
T_2(Q^2,s,u) p_\mu q_\nu + T_3(Q^2,s,u) q_\mu p_\nu 
\nonumber
\\
+ T_4(Q^2,s,u)q_\mu q_\nu + T_5(Q^2,s,u)g_{\mu\nu}
\Big \}
\label{2:Ti}
\ea
with
\ba
\lefteqn{T_1(Q^2,s,u)=\frac{2u\left[e_1\varphi_P(u)-
e_2\varphi_P(\bar{u})\right]}{\bar{u}Q^2-us+\bar{u}u m_P^2}}
\nonumber
\\
&&+\frac{1}{(\bar{u}Q^2-us+\bar{u}u m_P^2)^2}\Bigg\{
\frac{4f_{3P}}{f_P}\int\limits^{u}{\cal D}\alpha_i
\left[e_1m_{q_1}\varphi_{3P}(\alpha_i)-
e_2m_{q_2}\varphi_{3P}(\bar{\alpha_i})\right]
\nonumber
\\
&&
-2u\Bigg(4\left[e_1g_{1P}(u)-e_2g_{1P}(\bar{u})\right]
-4\left[e_1G_{2P}(u)-e_2G_{2P}(\bar{u})\right]-
2u\left[e_1g_{2P}(u)+e_2g_{2P}(\bar{u})\right]
\nonumber
\\
&&+\int\limits^{u}{\cal D}\alpha_i
\Big[(1-2v)\left(2[e_1\varphi_{\perp P}(\alpha_i)+
e_2\varphi_{\perp P}(\bar{\alpha_i})]
+\left[e_1\varphi_{\parallel P}(\alpha_i)+e_2\varphi_{\parallel P}(\bar{\alpha_i})\right] \right)
\nonumber
\\
&&
-2[e_1\widetilde{\varphi}_{\perp P}(\alpha_i)
-e_2\widetilde{\varphi}_{\perp P}(\bar{\alpha_i})]
-[e_1\widetilde{\varphi}_{\parallel P}(\alpha_i) 
-e_2\widetilde{\varphi}_{\parallel P}(\bar{\alpha_i})] 
\Big]\Bigg)\Bigg\}\,,
\label{T1}
\ea
\ba
\lefteqn{
T_2(Q^2,s,u)= -\frac{e_1\varphi_P(u)-e_2\varphi_P(\bar{u})}{\bar{u}Q^2
-us+\bar{u}u m_P^2}}
\nonumber
\\
&&
+\frac{1}{(\bar{u}Q^2-us+\bar{u}u m_P^2)^2}\Bigg\{
\frac{\mu_P}{3}\left[e_1m_{q_1}\varphi_{\sigma P}(u)-
e_2m_{q_2}\varphi_{\sigma P}(\bar{u})\right]
\nonumber
\\
&&+4\left[ e_1g_{1P}(u)-e_2g_{1P}(\bar{u})\right]
- 4\left[ e_1G_{2P}(u)-e_2G_{2P}(\bar{u})\right]- 
4u\left[e_1 g_{2P}(u)+e_2g_{2P}(\bar{u})\right]
\nonumber
\\
&&
+\int\limits^{u}{\cal D}\alpha_i
\Bigg[4(1-v)\left[e_1\varphi_{\perp P}(\alpha_i)
+e_2\varphi_{\perp P}(\bar{\alpha_i})\right]+
(1-2v)\left[e_1\varphi_{\parallel P}(\alpha_i)
+e_2\varphi_{\parallel P}(\bar{\alpha_i})\right]
\nonumber
\\
&&
-4(1-v)\left[e_1\widetilde{\varphi}_{\perp P}(\alpha_i)
-e_2\widetilde{\varphi}_{\perp P}(\bar{\alpha_i})\right]-
\left[e_1\widetilde{\varphi}_{\parallel P}(\alpha_i)
-e_2\widetilde{\varphi}_{\parallel P}(\bar{\alpha_i})
\right] \Bigg]\Bigg\}\,,
\label{T2}
\ea
\ba
\lefteqn{T_3(Q^2,s,u)= -\frac{e_1\varphi_P(u)-e_2\varphi_P(\bar{u})}{\bar{u}Q^2-us+\bar{u}u m_P^2}}
\nonumber
\\
&&
+\frac{1}{(\bar{u}Q^2-us+\bar{u}u m_P^2)^2}\Bigg\{
\frac{-\mu_P}{3}\left[e_1m_{q_1}\varphi_{\sigma P}(u)
-e_2m_{q_2}\varphi_{\sigma P}(\bar{u})\right]
\nonumber
\\
&&
+4\left[e_1g_{1P}(u)-e_2g_{1P}(\bar{u})\right]
- 4\left[e_1G_{2P}(u)-e_2G_{2P}(\bar{u})\right]
- 4u\left[e_1g_{2P}(u)+e_2g_{2P}(\bar{u})\right]
\nonumber
\\
&&+
\int\limits^{u}{\cal D}\alpha_i
\Bigg[-4v\left[e_1\varphi_{\perp P}(\alpha_i)+
e_2\varphi_{\perp P}(\bar{\alpha_i})\right]
+(1-2v)\left[e_1\varphi_{\parallel P}(\alpha_i)+
e_2\varphi_{\parallel P}(\bar{\alpha_i})\right]
\nonumber
\\
&&-4v\left[e_1\widetilde{\varphi}_{\perp P}(\alpha_i)
-e_2\widetilde{\varphi}_{\perp P}(\bar{\alpha_i})\right]-
\left[
e_1\widetilde{\varphi}_{\parallel P}(\alpha_i)
-e_2\widetilde{\varphi}_{\parallel P}(\bar{\alpha_i})\right] 
\Bigg]\Bigg\}\,,
\label{T3}
\ea
\ba
&&T_4(Q^2,s,u)= 4\frac{\left[e_1g_{2P}(u)+
e_2g_{2P}(\bar{u})\right]}{(\bar{u}Q^2-us+\bar{u}u m_P^2)^2}\,,
\label{T4}
\ea
\ba
\lefteqn{ T_5(Q^2,s,u)=-\frac{Q^2+s+(u-\bar{u})m_P^2
}{2(\bar{u}Q^2-us+\bar{u}u m_P^2)}
\left[e_1\varphi_P(u)-e_2\varphi_P(\bar{u})\right]}
\nonumber
\\
&&
+\frac{\mu_P}{(\bar{u}Q^2-us+\bar{u}u m_P^2)}
\left[e_1m_{q_1}\varphi_{p P}(u)-e_2m_{q_2}\varphi_{p P}(\bar{u})\right]
\nonumber
\\
&&+\frac{Q^2+s+(u-\bar{u})m_P^2
}{2(\bar{u}Q^2-us+\bar{u}u m_P^2)^2}
\Bigg\{4\left[e_1g_{1P}(u)-e_2g_{1P}(\bar{u})\right]
-4\left[e_1 G_{2P}(u)-e_2G_{2P}(\bar{u}) \right]
\nonumber
\\ 
&&+\int\limits^{u}{\cal D}\alpha_i
\Bigg[(1-2v)\left[e_1\varphi_{\parallel P}(\alpha_i)+
e_2\varphi_{\parallel P}(\bar{\alpha_i})\right]-
\left[e_1\widetilde{\varphi}_{\parallel P}(\alpha_i)
-e_2\widetilde{\varphi}_{\parallel P}(\bar{\alpha_i})\right] \Bigg]
\Bigg\} 
\,,
\label{T5}
\ea
where $s=(p-q)^2$; $\bar{u}=1-u$; $\alpha_i=\alpha_1,\alpha_2,
1-\alpha_1-\alpha_2$; $\bar{\alpha}_i= \alpha_2,\alpha_1,1-\alpha_1-\alpha_2$ 
and 
$$
\int\limits^{u}{\cal D}\alpha_i\equiv 
\int\limits_0^u d\alpha_1
\int\limits_0^{1-u}\frac{d\alpha_2}{1-\alpha_1-\alpha_2},~
v=\frac{u-\alpha_1}{1-\alpha_1-\alpha_2}\,. 
$$
In the above, $\varphi_P$ is a generic notation for the 
twist 2 DA of a pseudoscalar meson 
$P=\pi$ or $K$, whereas $\varphi_{pP}$, $\varphi_{\sigma P}$,
$\varphi_{3P}$ and $g_{1P}$, $g_{2P}$,  $\varphi_{\perp P,~\parallel P}$,
$\tilde{\varphi}_{\perp P,~\parallel P}$ are, respectively, DA of twist 
3 and 4. Their definitions taken from Ref.~\cite{BF} 
(see also Ref.~\cite{Ball})  
are collected in App.~\ref{AppB}. The decay constant $f_P$ of $P$ is 
defined as
$\langle 0 \mid \bar{q}_2\gamma_\mu \gamma_5 q_1\mid P(p)\rangle=if_Pp_\mu$.
Furthermore, $\mu_P=m_P^2/(m_{q_1}+m_{q_2})$ is 
the twist 3 DA normalization factor and
$
G_{2P}(u)=\int_0^u dv g_{2P}(v)\,.
$  
In the case of nonstrange quarks, $q_1=q_1'=u$, $q_2=q_2'=d$, both chiral 
and isospin symmetry limits can safely be adopted.
In this limit the $u,d$ quark 
masses as well as the pion mass are neglected and
DA are either symmetric or antisymmetric 
(see App.~\ref{AppB}) \footnote{The type of symmetry is established
applying $G$-parity transformation to the underlying matrix elements.},
with respect to the replacements 
$u \leftrightarrow \bar{u}$, or  $\alpha_1\leftrightarrow
\alpha_2$. In this  case the twist 3 parts in 
Eqs.~(\ref{T1})--(\ref{T5}) vanish and the 
combination of quark charges 
$e_1-e_2=e_u-e_d=1$ factorizes out. The resulting expression for $T_1$
coincides with the one obtained in Ref.~\cite{BH} 
except that the signs of the terms containing the 
twist 4 quark-gluon DA $\varphi_{\perp P,\,\parallel P}$
are opposite.  The same discrepancy in signs  
is found in the expressions for $T_i$ obtained  
in the chiral limit in Ref.~\cite{Belyaev}
comparing them with  Eqs.~(\ref{T1})-(\ref{T5}). 
In the next section we will demonstrate that Eqs.~(\ref{T1})-(\ref{T5})
are fully consistent with the relations obtained from QCD equations of 
motion. Finally, we note that the twist 3 terms 
in Eqs.~(\ref{T1})-(\ref{T5}) are new.

\section{Ward Identities}
\label{Sect3}

Multiplying the correlation function (\ref{2:cor})
by the four-momentum $q$ one obtains
\ba
\label{WIvector}
q^\nu T_{\mu\nu} = -\int\! d^4 x \,e^{iqx}
\Bigg(\langle 0| T\{\bar{q}_2(0)\gamma_\mu\gamma_5 q_1(0)
\frac{\partial}{\partial x_\nu}\Big( 
e_1\bar{q}_1(x)\gamma_\nu q_1'(x)
+e_2\bar{q}_2'(x)\gamma_\nu q_2(x)\Big)\}
|P(p)\rangle 
\nonumber
\\
-\delta(x_0) \langle 0|
[\bar{q}_2(0)\gamma_\mu\gamma_5 q_1(0), (e_1\bar{q}_1(0,\vec{x})\gamma_0 
q_1'(0,\vec x)+e_2\bar{q}_2'(0,\vec x)\gamma_0 q_2(0,\vec x))]
|P(p)\rangle\Bigg )
\;,
\ea
where the second term containing equal-time current 
commutators originates from the differentiation 
of the $\theta(x_0)$ in the T-product of currents.
For the conserved vector currents $q_1=q_1'$ and $q_2=q_2'$ 
the first term on the r.h.s. of Eq.~(\ref{2:cor}) vanishes.
For the second term the standard commutation relations for the 
equal-time current densities can be employed, e.g., in the case of 
the pion:
\ba
\Big[\bar{d}(0)\gamma_\mu\gamma_5 u(0), \Big(e_u\bar{u}(0,\vec{x})\gamma_0
u(0,\vec x)+e_d\bar{d}(0,\vec{x})\gamma_0 d(0,\vec{x})\Big)\Big]=
\delta^{(3)}(\vec x) \bar{d}(0,\vec{x})\gamma_\mu\gamma_5 u(0,\vec{x})\,,
\ea
yielding for the correlation function the Ward identity
\ba
q^\nu T_{\mu\nu}=i f_\pi p_\mu\,.
\label{WI}
\ea
In the chiral limit $m_{q_1}=m_{q_2}=0$ the 
axial-vector current 
is also conserved. Hence,  we get an additional relation
\ba
\label{WIaxial}
(p-q)^\mu T_{\mu\nu}^{} = -i f_\pi p_\nu\,.
\ea
The above Ward identities are valid for arbitrary $q$ and $p$.
This circumstance allows one
to get relations between various pion DA by substituting Eq.~(\ref{2:Ti})
in l.h.s. of Eqs.~(\ref{WI}),(\ref{WIaxial}).

Here we will only concentrate on the  chiral limit, so that 
both Eqs. (\ref{WI}) and (\ref{WIaxial}) are valid and $p^2=0$.
It is easy to check that the r.h.s. of these equations are saturated by 
the twist-2 contribution to their l.h.s. 
Hence, the Ward identities (\ref{WI}) and (\ref{WIaxial}) 
yield nontrivial relations between two- and three-particle DA of twist
4. Note that in the chiral limit 
different twists are separated by dimensions, therefore 
contributions to the correlation function with twist higher than 4 neglected 
in our calculation are unimportant\footnote{
In fact, we also neglect four-particle Fock components of twist 4
in the light-cone expansion of the matrix elements. This is 
consistent with the approximation adopted in deriving the relations 
from QCD equations of motion \cite{BF}.}. 
Using
\be
\frac{2q.p}{(q-up)^4} = \frac{\partial}{\partial u}\frac{1}{(q-up)^2}
\quad\mbox{and}\quad
\frac{q^2}{(q-up)^4} = 
\frac{1}{(q-up)^2}+u\frac{\partial}{\partial u}\frac{1}{(q-up)^2}\,,
\ee
together with partial integration in $u$,
rewriting all twist 4 contributions in Eqs. (\ref{WI}) and
(\ref{WIaxial}) as $\int_0^1du\frac{1}{(p-uq)^2}F(u)$
and then extracting $F(u)=0$, one obtains the following relations:
\ba
g_{2\pi}(u) &=&\int\limits^{u}{\cal D}\alpha_i
(\left(\varphi_{\perp\pi}(\alpha_i)-(1-2v)
 \widetilde\varphi_{\perp\pi}(\alpha_i)\right)\;,
\label{WIA}
\ea
\ba
G_{2\pi}(u) &=& \frac{u}{2}g_{2\pi}(u)-\frac{1}{2}\int\limits^{u}{\cal D}\alpha_iv
\left(\varphi_{\perp\pi}(\alpha_i)+\widetilde\varphi_{\perp\pi}(\alpha_i)\right)\;,
\label{WIB}
\ea
\ba
g_{1\pi}(u) &=& G_{2\pi}(u)+\frac{1}{2}u\bar{u}g_{2\pi}^\prime(u)
\nonumber\\&&-\frac{1}{4}\int\limits^{u}{\cal D}\alpha_i
\left[
(1-2v)\left(\varphi_{\parallel\pi}(\alpha_i)+2\varphi_{\perp\pi}(\alpha_i)\right)
-\widetilde\varphi_{\parallel\pi}(\alpha_i)-2\widetilde\varphi_{\perp\pi}(\alpha_i)
\right]\,,
\label{WIC}
\ea
where $g_{2\pi}^\prime(u)=\partial g_2(u)/\partial u$.
We notice that the above expressions can be used to rewrite in the chiral limit
($P=\pi$) the twist 4 part of the correlation function (\ref{2:Ti})
using only one DA $g_{2\pi}$ and its derivative over $u$, so that
\ba 
T_{\mu\nu} = if_\pi\int_0^1 du \Bigg\{
\frac{1}{\bar{u}Q^2-us}\Big(2up_\mu p_\nu-q_\mu p_\nu-p_\mu q_\nu-
(q\cdot p)g_{\mu\nu}\Big)\varphi_\pi(u)
\nonumber\\
+\frac{2}{(\bar{u}Q^2-us)^2}
\Bigg\{p_\mu p_\nu \Big(2u^2g_{2\pi}(u)-2u^2\bar u g_{2\pi}^\prime(u)\Big)
+(p_\mu q_\nu+q_\mu p_\nu)(-2ug_{2\pi}(u)+u\bar u g_{2\pi}^\prime(u))
\nonumber\\
+(p_\mu q_\nu-q_\mu p_\nu+2q_\mu q_\nu)g_{2\pi}(u)
+g_{\mu\nu}\left[\left (2Q^2+(q.p)(1+2u)\right) g_{2\pi}(u)
- (q.p)\, u\bar u g_{2\pi}^\prime(u)  \right]\Bigg\}\,.
\label{finalT}
\ea

The relations (\ref{WIA}) and (\ref{WIB}) can also be obtained 
using the technique of QCD equations of motion \cite{BF}. 
The starting objects in this case 
are the derivatives of quark-antiquark operators expressed via 
quark-antiquark-gluon operators, e.g.:
\be
\label{BFrel}
\frac{\partial}{\partial x_\mu}
\left\langle 0|\overline{d}(0)\gamma_\mu\gamma_5 u(x)|\pi^+(p)
\right\rangle = i\!\int_0^1\! \alpha d\alpha \left\langle 0|
\overline{d}(0)\gamma_\mu\gamma_5 x_\lambda G^{\lambda \mu}(\alpha x)u(x)
|\pi^+(p) \right\rangle
\ee
and 
\be
\label{Bijrel}
\frac{\partial}{\partial x_\nu}
\left\langle 0|\overline{d}(0)\gamma_\mu \gamma_\beta\gamma_\nu\gamma_5
 u(x)| \pi^+(p)\right\rangle
=i\!\int_0^1 \!\alpha d\alpha \left\langle 0|
\overline{d}(0)\gamma_\mu\gamma_\beta\gamma_\nu\gamma_5 
x_\lambda G^{\lambda \nu}(\alpha x)u(x)
|\pi^+(p) \right\rangle,
\ee
where $G_{\mu\nu}=g_sG^a_{\mu\nu}(\lambda_a/2)$,
$tr\lambda_a\lambda_b=2\delta^{ab}$.
The relations derived from Eq.~(\ref{BFrel}) are
\ba
g_{1\pi}(u) &=& \frac{u}{2}g_{2\pi}(u)-G_{2\pi}(u)+\frac{1}{2}
\int\limits^{u}{\cal D}\alpha_i
v\left(\varphi_{\parallel\pi}(\alpha_i)-2\varphi_{\perp\pi}(\alpha_i)\right)\,,
\label{bf1}
\ea
and its $u\leftrightarrow \bar{u}$ equivalent:
\ba
g_{1\pi}(u) &=& -\frac{\bar{u}}{2}g_{2\pi}(u)-G_{2\pi}(u)-\frac{1}{2}
\int\limits^{u}{\cal D}\alpha_i
(1-v)\left(\varphi_{\parallel\pi}(\alpha_i)-
2\varphi_{\perp\pi}(\alpha_i)\right)\;.
\label{bf2}
\ea
Combining Eqs.~(\ref{bf1}) and (\ref{bf2}) 
one gets the two relations obtained in Ref.~\cite{BF}.
Eq.~(\ref{Bijrel}) which was also used in Ref.~\cite{Belyaev} yields
\ba
g_{1\pi}(u) &=& -\frac{u}{2}g_{2\pi}(u)+G_{2\pi}(u)+\frac{1}{2}
\int\limits^{u}{\cal D}\alpha_i
v\left(\varphi_{\parallel\pi}(\alpha_i)+2\widetilde\varphi_{\perp\pi}(\alpha_i)\right)\;,
\label{Bel}
\ea
and 
\ba
g_{1\pi}(u) &=& \frac{\bar{u}}{2}g_{2\pi}(u)+G_{2\pi}(u)+\frac{1}{2}
\int\limits^{u}{\cal D}\alpha_i
(1-v)\left(-\varphi_{\parallel\pi}(\alpha_i)+2\widetilde\varphi_{\perp\pi}(\alpha_i)
\right)\,.
\label{Bij}
\ea
Combining Eq.~(\ref{Bel}) and Eq.~(\ref{Bij})
with Eqs.~(\ref{bf1}) and (\ref{bf2}), after simple
algebra one  indeed reproduces the relations obtained from Ward identities,
but only  two of them, Eqs.~(\ref{WIA}) and (\ref{WIB}).
The relation (\ref{WIC}), the only one involving
the DA $\widetilde\varphi_{\parallel \pi}$ is new and was not obtained 
using equations of motion.
Note that the observed consistency between the relations derived from Ward
identities and from QCD equations of motion provides an independent 
check of our result for the correlation function. Indeed,
taking the correlation function calculated in Ref.~\cite{Belyaev}
with different signs at the terms with $\varphi_{\perp\pi,\parallel
\pi}$ we obtain a contradiction between the two types of relations.

If the  chiral symmetry is violated, $m_q \sim m_P^2\neq 0$, 
the Ward identity (\ref{WI}) based on the conservation 
of the e.m. current is still valid, but the relations 
following from this identity are more complicated, 
mixing DA of twist 2,3 and 4 and not allowing  
to reduce the twist 4 part to an integral over 
a single DA.
The corresponding analysis goes beyond the scope of this paper. 

\section{Updated prediction for the pion e.m. form factor}
\label{Sect4}

The LCSR for the pion e.m. form factor was originally derived in
Ref.~\cite{BH}. Let us briefly outline the procedure. 
The part of the correlation function (\ref{2:Ti})
(in the chiral limit) proportional to $ \sim p_\mu p_\nu $ 
was matched to the hadronic dispersion relation 
in the variable $s=(p-q)^2$, that is, in the channel of the axial-vector current:
\be
if_\pi\int\limits_0^1 du T_1(Q^2,s,u) = 
\frac{2if_\pi F_\pi(Q^2)}{-s}+ 
\int\limits_{s_0^h}^\infty \frac{\rho^h(s')ds'}{s'-s}\,.
\label{disp1}
\ee
In this relation, the first term on the r.h.s. is the ground-state
contribution of the pion where $f_\pi=132 $ MeV and 
$F_\pi(Q^2)$ is the pion e.m.
form factor defined in the standard way:
\be
F_\pi(Q^2)(2p-q)_\nu =\langle\pi^+(p-q)\mid j^{em}_\nu \mid \pi^+(p)\rangle\,,
\ee
$j^{em}_\nu$ being the quark e.m. current.
The contributions of the $a_1$ meson and excited states
with $J^P=O^-,1^+$ form the spectral
density $\rho^h$ which is estimated as usual, with the help
of quark-hadron duality:
\be
\rho^h(s) \Theta(s-s_0^h)=\frac{if_\pi}{\pi} \int\limits_0^1 du\, 
\mbox{Im}_s T_1(Q^2,s,u)\Theta(s-s_0^\pi)\,,
\label{dual}
\ee
where the effective threshold
parameter $s_0^{\pi}=0.7$ GeV$^2$ is determined from the
SVZ sum rule \cite{SVZ} for the correlator of two 
$\bar{u}\gamma_\mu\gamma_5d$ currents.
Representing the l.h.s of Eq.~(\ref{disp1}) in a form of the 
dispersion integral: 
\be
\int\limits_0^1 du\, T_1(Q^2,s,u)=\frac1\pi
\!\int\limits_0^{\infty}\! \frac{ds'}{s'-s} 
\int\limits_0^1\!du\, \mbox{Im}_{s'} T_1(Q^2,s',u)\,,
\label{disp2}
\ee
using Eq.~(\ref{dual}) and performing the Borel transformation ,
$(p-q)^2\to M^2$, we obtain the resulting sum rule:
\be
F_\pi(Q^2) =\frac{1}{2\pi}\int\limits_{0}^{s_0^{\pi}}\!ds\, e^{-s/M^2}
\!\int\limits_{0}^1du~\mbox{Im}_s T_1(Q^2,s,u)\,.
\label{lcsr}
\ee
In the twist-2 approximation one has \cite{BH}: 
\be
F_\pi^{(2)}(Q^2)=\int_{u_0^\pi}^1 du~\varphi_\pi(u,\mu)
\exp\left( -\frac{\bar{u}Q^2}{uM^2}\right)\,.
\label{LCSRpitw2}
\ee
In the above $u_0^\pi= Q^2/(s_0^{\pi}+Q^2)$ and we have indicated
the dependence of the DA $\varphi_\pi$
on the normalization scale $\mu$.

The LCSR  (\ref{lcsr}) was further improved in
Ref.~\cite{BKM} where the
$O(\alpha_s)$ contribution to the twist 2 part
was calculated by taking into account the perturbative gluon 
exchanges between the quark lines in the diagram of Fig.~\ref{fig1}a.
For convenience we present in App.~\ref{AppC} the explicit expression
for $F_\pi^{(2,\alpha_s)}(Q^2)$ 
which has to be added to the r.h.s. of Eq.~(\ref{LCSRpitw2}).
Recall that this contribution provides the  
$\sim \alpha_s/Q^2$ asymptotic behavior \cite{exclusive} of the form
factor. As explained in detail in Ref.~\cite{BKM} the  form factor
obtained from LCSR includes both  
the hard-scattering and soft (end-point) contributions.

Our main update of the sum rule for $F_\pi(Q^2)$ concerns
the twist 4 term for which a new, corrected expression is 
obtained from  Eq.~(\ref{finalT}):
\begin{equation}
F_\pi^{(4)}(Q^2)= \int\limits ^1_{u_0^\pi}du \frac{\varphi^{(4)}_\pi(u,\mu)}{uM^2}
\exp\left( -\frac{\bar{u}Q^2}{uM^2}\right) + \frac{u_0^\pi\varphi^{(4)}_\pi
  (u_0^\pi,\mu)
}{Q^2}e^{-s_0^\pi/M^2}\,,
\label{tw4res}
\end{equation}
where 
\be
\varphi^{(4)}_\pi(u,\mu)= 2u\left(g_{2\pi}(u,\mu)-
\bar{u}g'_{2\pi}(u,\mu)\right)\,.
\label{phi4}
\ee

The second term on the r.h.s. of Eq.~(\ref{tw4res}) is a `surface term'
originating after the continuum  subtraction as
explained in Ref.~\cite{BKM}. In the same paper the factorizable 
twist 6 contribution to LCSR was calculated:  
\be
F_\pi^{(6)}(Q^2)=\frac{4\pi\alpha_sC_F}{3f_\pi^2Q^4}
\langle 0\!\mid\bar{q}q \mid \!0\rangle^2 
\ee
in terms of the quark condensate density (see Ref.~\cite{BKM} for 
the diagrams and other details concerning this contribution).
Note that the twist 6 term is numerically very
small starting from $Q^2 = 1$~GeV$^2$ which is 
therefore a natural lower boundary of the LCSR 
validity region~\footnote{Recent work on the pion and kaon form factors 
at low momentum transfers can be found in Ref.~\cite{lowenergy}.}.

We turn now to the numerical calculation of the pion form factor,
\be
F_\pi(Q^2)=F_\pi^{(2)}(Q^2)+ F_\pi^{(2,\alpha_s)}(Q^2)+
F_\pi^{(4)}(Q^2)+F_\pi^{(6)}(Q^2)\,,
\label{contrib}
\ee
where twist 2,4 and the factorizable twist-6 contributions to LCSR
are added together. In our numerical evaluation of Eq.~(\ref{contrib}), 
following Ref.~\cite{BKM} we take $ 0.8 < M^2 < 1.5 $ GeV$^2$ and adopt the variable
normalization scale $\mu^2_u =(1-u)Q^2+u M^2$
of the light-cone DA. The same scale is adopted for the 
normalization of $\alpha_s$. For the latter  
the two-loop running is used with $\bar{\Lambda}^{(3)}=340$ MeV 
corresponding to $\alpha_s(\mbox{1 GeV})=0.48$. For the twist 2 pion DA  
we take the asymptotic form $\varphi_\pi(u)=6u(1-u)$.  
The influence of nonasymptotic corrections will be discussed later.
Concerning  the twist 4 pion DA we actually
need only one of them, $g_{2\pi}$. Interestingly,
in first order of the conformal expansion \cite{BF} 
this DA does not contain nonasymptotic contributions.
Using the asymptotic form of $g_{2\pi}(u)$ presented in App.~\ref{AppB}
one obtains a compact expression:
\be
\varphi^{(4)}_\pi(u,\mu)=\frac{20}3\delta^2_\pi(\mu)u\bar{u}(1-u(7-8u))\,.
\label{phi41}
\ee
The nonperturbative parameter 
$\delta^2_\pi\simeq $ 0.2 GeV$^2$
determining the vacuum-to-pion matrix element of the 
quark-antiquark-gluon current (see definition in App.~\ref{AppB}) 
was estimated from various 2-point QCD
sum rules in Ref.~\cite{delta}. To assess the 
theoretical uncertainty, we have recalculated $\delta^2_\pi$ 
using 
the diagonal sum rule for two quark-antiquark-gluon currents
which is less dependent on the variations of quark and gluon condensates.
The result, in agreement with Ref.~\cite{delta}, is
\be
\delta^2_\pi( 1 \mbox{GeV})=0.17 \pm 0.05~\mbox{GeV}^2\, 
\label{delta}
\ee
obtained with 
$ \langle 0\!\!\mid \!\bar{q}q \!\mid \!\!0\rangle=
(-240\pm 10~\mbox{MeV})^3$ 
and $\langle 0 \!\!\mid \!(\alpha_s/\pi)G^a_{\mu\nu}G^{a\mu\nu}\!\mid \!\!0 \rangle
=0.012 \pm 0.006~\mbox{GeV}^4$. 

\begin{figure}[t]
\begin{center}
\includegraphics[height=0.6\textwidth]{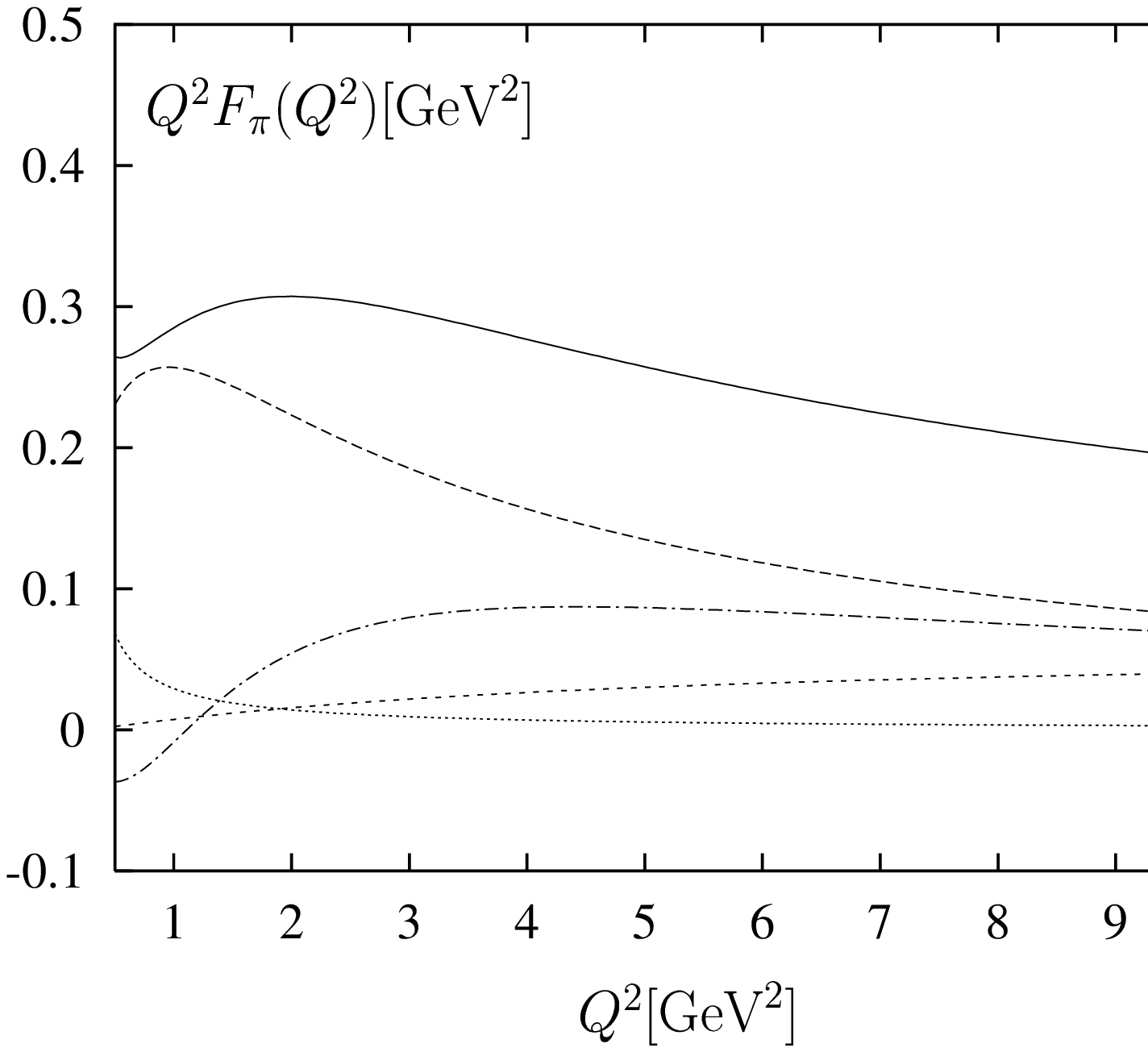}
\end{center}
\caption{ {\it Pion e.m. form factor obtained from LCSR
with the asymptotic pion DA (solid) including  the 
twist 2 leading-order (long-dashed), twist 2 $O(\alpha_s)$
(short-dashed), twist 4 (dash-dotted) and 
factorizable twist 6  (dotted) contributions.}
\label{fig2}}
\end{figure}

Our prediction for the pion e.m. form factor given by  Eq.~(\ref{contrib}) is 
plotted in Fig.~\ref{fig2}, calculated with the 
asymptotic pion DA at a typical value 
of $ M^2=1$ GeV$^2$, and at 
$\mu=\mu_u$ and $\delta^2_\pi= 0.17~\mbox{GeV}^2$.
Importantly, the corrected
twist 4 contribution is about two times larger than 
estimated before \cite{BH,BKM}. Note that at $Q^2\to \infty$
the twist 4 term given by Eq.~(\ref{tw4res})
has the same $\sim 1/Q^4$ asymptotic behavior
as the twist 2 contribution (\ref{LCSRpitw2})
\footnote{Contributions nonvanishing or growing  
with  $Q^2$ are absent in LCSR. Such anomalous contributions 
emerge 
in QCD sum rules based on the local condensate expansion
\cite{IofSmil,Rad82} making the latter not applicable at 
$Q^2 \gg 1$ GeV$^2$.},
but has one extra power of $1/M^2$. This can be seen
explicitly by rewriting Eq.~(\ref{tw4res}), with the help of 
Eq.~(\ref{phi41}), in the form of a dispersion integral 
with the integration variable $s=Q^2\bar{u}/u$:
\be
\label{eq:38}
F_\pi^{(4)}(Q^2)=\frac{40}{3}\delta^2(\mu)\int\limits_0^{s_0^\pi}
ds e^{-s/M^2}\frac{Q^8}{(Q^2+s)^6}\left( 1-\frac{9s}{Q^2}+\frac{9s^2}{Q^4}-
\frac{s^3}{Q^6}\right)\,,
\ee
yielding at $Q^2 \to \infty$  
\be
F_\pi^{(4)}(Q^2) \sim \frac{40\delta^2 M^2}{3 Q^4}\left(1-e^{-s_0^\pi/M^2}\right)\,,
\label{tw4as}
\ee
to be compared with the corresponding limit of 
Eq.~(\ref{LCSRpitw2}):
\be
F_\pi^{(2)}(Q^2) \sim \frac{6M^4}{Q^4}\left(1-\left(1+\frac{s_0^\pi}{M^2}
\right)e^{-s_0^\pi/M^2}\right)\,.
\label{tw2as}
\ee
Although the twist 4 term has indeed an extra suppression 
factor $\delta^2/M^2$ as compared with the twist 2 term, the 
overall numerical coefficients in Eqs.~(\ref{tw4as}) and (\ref{tw2as})
are of the same order at $M^2\sim $ 1 GeV$^2$.

\begin{figure}[t]
\begin{center}
\includegraphics[width=0.7\textwidth]{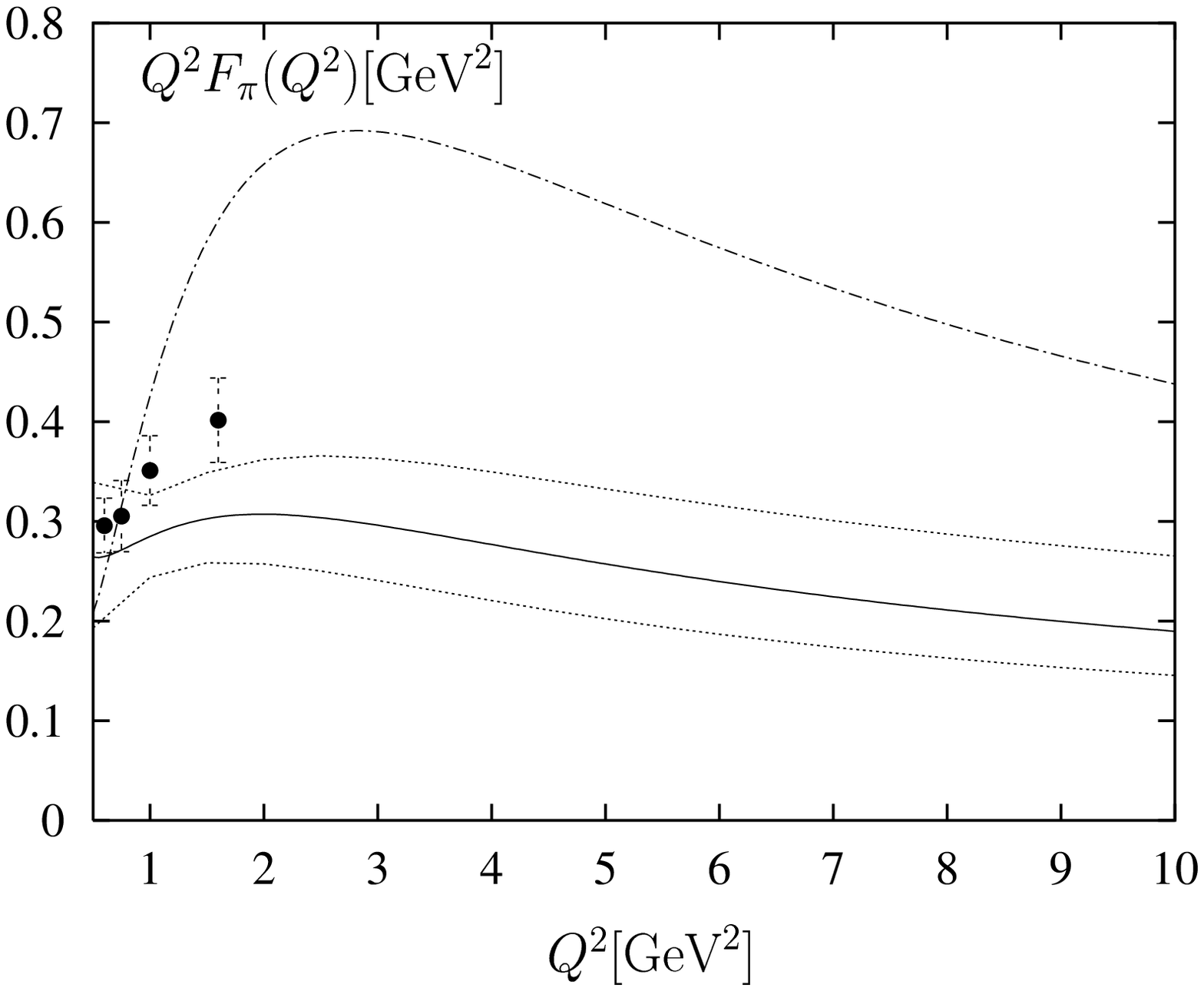}
\end{center}
\caption{ {\it The pion e.m. form factor calculated from 
LCSR in comparison with the 
Jefferson Lab data \cite{Volmer} shown with points
(the experimental error and the model uncertainty 
are added in quadratures). 
The solid line corresponds to the asymptotic 
pion DA, dashed  lines indicate the estimated 
overall theoretical uncertainty ; the dash-dotted
line is calculated with the CZ model \cite{CZ84}
of the pion DA. }
\label{fig3}}
\end{figure}

The LCSR approach allows one to estimate 
the theoretical uncertainty of the predicted form factor. 
We did it in the following way. 
First of all, $M^2$ and $\delta^2_\pi$  
were varied within allowed intervals. 
Furthermore, in order to investigate the sensitivity to the 
choice of the renormalization scale 
our calculation was repeated at two fixed scales $Q^2$ and 
$M^2$ adopting the variation of the results as a theoretical uncertainty.
Finally, accounting for the missing twist $\geq $6 terms we 
assume that the absence of the latter introduces 
an additional uncertainty equal to $\pm F_\pi^{(6)}(Q^2)$. 
All abovementioned  variations of the LCSR prediction for $F_\pi(Q^2)$ 
are then added linearly which is a rather conservative approach.

In Fig.~\ref{fig3} we plot $F_\pi(Q^2)$, 
calculated with 
the asymptotic pion DA and at $M^2=1$ GeV$^2$. 
The resulting uncertainty of the form factor indicated in this figure is about 
$\pm (20\div 30)\%$ at $Q^2 \geq 1 GeV^2$. 
At  $Q^2 <$ 1 GeV$^2$ the uncertainty grows revealing the 
inapplicability of LCSR for small momentum transfers. 
In the region $2.0\leq Q^2 <10$ GeV$^2$ our prediction for the 
pion e.m. form factor with the asymptotic pion DA can be fitted
to the following simple formula:
\be
Q^2F_\pi(Q^2)= (0.0735\div 0.2016)+\frac{0.7908\div 0.9340}{Q^2}-\frac{0.8496\div1.2068}{Q^4},
\ee
(all numbers in GeV$^2$),
where the correlated intervals take into account 
the total uncertainty.

The accuracy of the LCSR prediction (\ref{contrib}) can be 
improved further by including 
various higher-twist corrections (due to twist-4 multiparticle and twist
6 DA) which were not yet analyzed. However, 
the smallness of the factorizable twist 6 term indicates that these effects 
are most probably numerically unimportant. In addition,
it is desirable to improve also the perturbative expansion
of the correlation function calculating  
the $O(\alpha_s)$ term of twist 4 and the  
$O(\alpha_s^2)$ term of  twist 2. An attempt to account for 
the latter was made in Ref.~\cite{BKM} by matching LCSR 
to the NLO perturbative calculation \cite{Melic}.

Finally, in Fig.~\ref{fig3} we compare  
our numerical prediction with the 
recent accurate data on $F_\pi(Q^2)$ obtained 
from the pion electroproduction 
at Jefferson Lab \cite{Volmer} at $Q^2= 0.6 \div 1.65 $ GeV$^2$,
in the region which only partly overlaps 
with the LCSR validity region $Q^2>1$ GeV$^2$.    
We find that within theoretical uncertainties and experimental errors
the form factor calculated with the asymptotic pion DA $\varphi_\pi(u)$
is consistent with data \footnote{
LCSR predictions also agree with older measurements 
of $F_\pi(Q^2)$ at $Q^2=1\div 6$ GeV$^2$ which however have  
large experimental errors.}. 

With $F_\pi(Q^2)$ accurately measured at the whole 
region $Q^2=1\div 10$ GeV$^2$  
it should in principle be possible to constrain/fit the nonasymptotic
part of $\varphi_\pi(u)$ determined  by the coefficients 
$a_n$ in the expansion over Gegenbauer polynomials (see App.~\ref{AppB}).
Taking into account the complete expansion one obtains
\be
F_\pi(Q^2)= [F_\pi(Q^2)]_{as} + 
\sum\limits_{n=1}^{\infty}a_{2n}(\mu_0)f_{2n}(Q^2,\mu,\mu_0)\,,
\label{fit}
\ee
where the first term on the r.h.s. is the form factor calculated 
with the asymptotic DA and in the sum each $a_{2n}$ is 
multiplied by a calculable function:
\be
f_{2n}(Q^2,\mu,\mu_0)=
6\left(\frac{\alpha_s(\mu)}{\alpha_s(\mu_0)}\right)^{-\gamma_n^{(0)}/\beta_0}
\int_{u_0^\pi}^1du~u \bar u C_{2n}^{3/2}(u - \bar u)
\exp\left( -\frac{\bar{u}Q^2}{uM^2}\right)+...\,.
\label{fn}
\ee
In the above, $\mu_0\sim $ 1 GeV is a certain low scale, 
and  the anomalous dimensions $\gamma_n$ of
the renormalization factors are given in Appendix \ref{AppB}.
For brevity, the $O(\alpha_s)$-correction to Eq.~(\ref{fn})
is denoted by ellipses. 

\begin{figure}[t]
\begin{center}
\includegraphics[width=0.7\textwidth,angle=0]{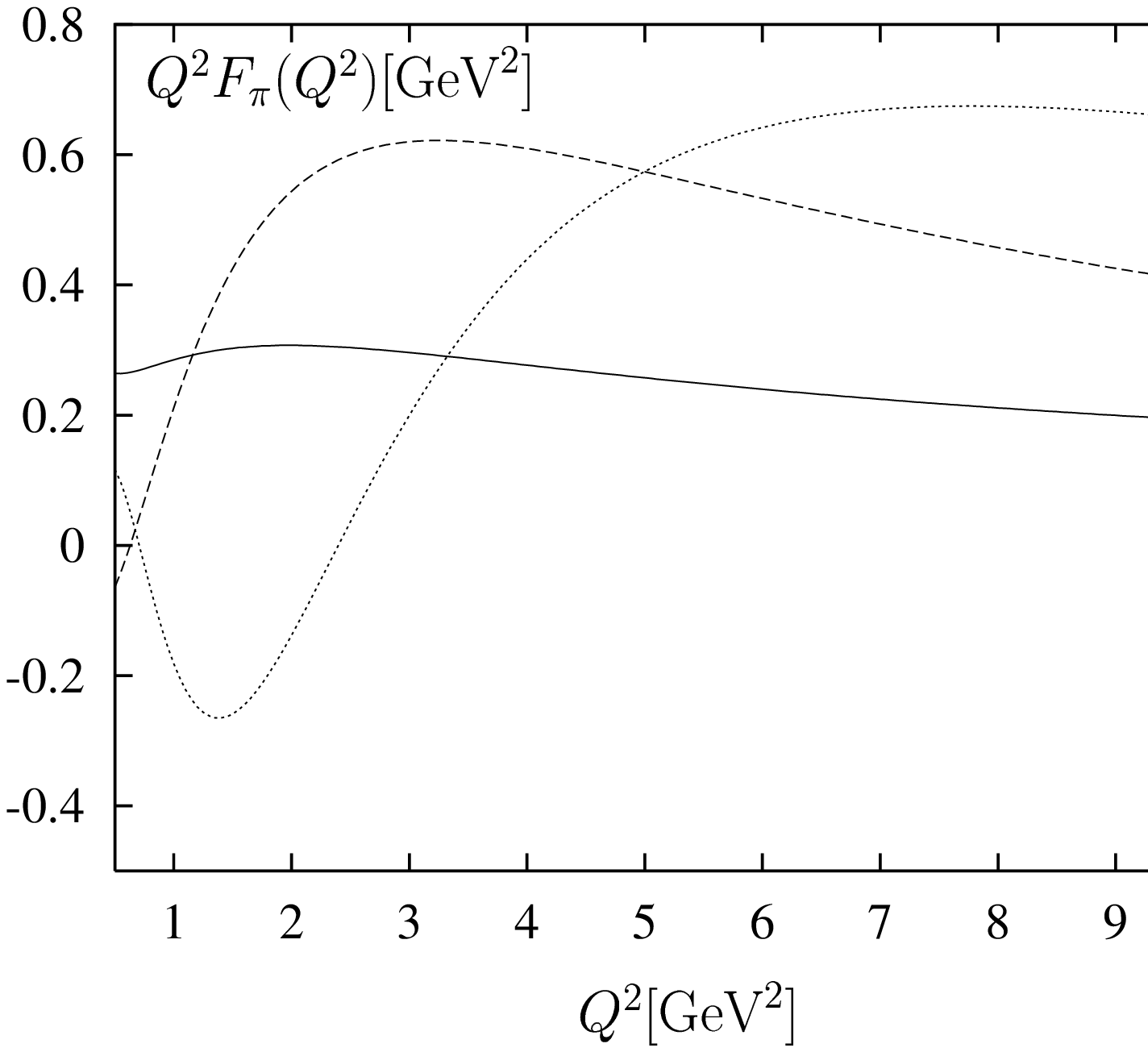}
\end{center}
\caption{ {\it Graphical illustration to Eq.(\ref{fit}). 
The pion e.m. form factor obtained from LCSR
with the asymptotic pion DA (solid line) and 
the coefficients at $a_2(1 GeV)$(long-dashed) and $a_4(1 GeV)$ 
(short-dashed). }
\label{fig4}}
\end{figure}

A direct fit of all $a_n$ from Eq.~(\ref{fit}) is of course not 
a realistic task. In fact, using the arguments of the conformal partial
wave expansion, one expects that the coefficients are decreasing 
with $n$, $a_{2n+2}<a_{2n}$. Based on these arguments, 
the form of $\varphi_\pi(u)$ usually  discussed in the literature involves
one ($a_2$) or two ($a_2,a_4$) nonzero coefficients neglecting the rest. 
Having adopted a certain simple ansatz for $\varphi_\pi(u)$ one 
is then able to constrain or even fit the coefficients from
Eq.~(\ref{fit}). However, the current data \cite{Volmer} are sufficient to 
constrain only the simplest ansatz with a single nonzero coefficient
$a_2$. This can be seen from Fig.~\ref{fig4}
where $f_2$ and $f_4$ in Eq.~(\ref{fn}) are plotted
in comparison with $F_\pi^{as}$. Due to 
different signs of $f_2$ and $f_4$ at $Q^2\leq 3$ GeV$^2$ 
it is difficult to distinguish the form of $\varphi_\pi$
with $a_2\neq 0$ from the one where both $a_2,a_4\neq 0$. 
E.g. one obtains equally good fits to the experimental 
data shown in Fig.~(\ref{fig3}) 
with $a_2(1\mbox{GeV})=0.05$, $a_4(1\mbox{GeV})=-0.30$ 
as with $a_2(1\mbox{GeV}) = 0.25, a_4(1\mbox{GeV}) = 0$.  
If we impose that all $a_{n> 2}=0$ in Eq.~(\ref{fit})
the coefficient $a_2$ can be fitted to the following 
interval consistent with zero:
\be
a_2(1\mbox{GeV})=0.24 \pm 0.14  \pm 0.08\,,  
\ee
where the first error reflects our estimated theoretical
uncertainty whereas the second one corresponds to the experimental errors.
One needs data at larger $Q^2$ to resolve more complicated 
patterns of nonasymptotic coefficients in $\varphi_\pi$.

\section{The kaon electromagnetic form factor }
\label{Sect5}

The LCSR for the charged kaon e.m. form factor can be
easily obtained from the correlation function (\ref{T1}),
substituting  $P=K^+$, $e_1=e_u=+2/3$, 
$e_2= e_s=-1/3$, $m_{q_1}=0$, $m_{q_2}=m_s$.
We will systematically retain all $O(m_s) \sim O(m_K^2)$ 
effects which are numerous, in general. At the 
purely kinematical level one has to account for 
$p^2=m_K^2$ in the 
correlation function. Furthermore, the s-quark propagator
produces a chirally non-invariant part proportional to 
$m_s$ which brings the twist 3 contribution into the 
game (the $m_s^2$  in the denominator of the quark propagator
is neglected, being a higher-order effect). Finally, in the 
light-cone DA there are 
SU(3)-flavour symmetry ($SU(3)_{fl}$) violating corrections of three
types. First, the normalization factors, determining 
the quark-antiquark vacuum-to kaon matrix elements 
in the local limit $x=0$ differ from the corresponding 
factors for the pionic matrix elements, e.g. 
$f_K \neq f_\pi$. Secondly, the nonasymptotic parts
of the kaon DA are asymmetric with respect 
to the interchange of quark and antiquark 
fields, with a  larger average momentum fraction of the strange quark. 
At the twist 2 level this effect manifests itself
in the nonvanishing odd coefficients 
of the Gegenbauer expansion (\ref{phipi}):
$a_1^K, a_3^K, ... \sim m_s \neq 0 $. 
For the higher twist DA the $SU(3)_{fl}$ violating 
asymmetries  were not studied 
yet, and in our numerical calculation we will 
neglect them. On the other hand, we will take into account 
the so-called meson-mass corrections to the twist 4 DA investigated 
and worked out in Ref.~\cite{Ball}. These
effects include the mixing of nonasymptotic 
parts of twist 2,3 and 4 DA beyond the chiral limit. 
The corresponding expressions are presented in App.~\ref{AppB}.     

Apart from $SU(3)_{fl}$ violating corrections listed above 
the derivation of the LCSR for the 
kaon e.m. form factor repeats the procedure for the pion
outlined in the previous section.
The result reads:
\be
F_K(Q^2)=F_K^{(2)}(Q^2)+ F_K^{(2,\alpha_s)}(Q^2)
+ F_K^{(3)}(Q^2) + F_K^{(4)}(Q^2)+F_K^{(6)}(Q^2)\,,
\label{contribK}
\ee
where the twist 2 contribution is  
\be
F_K^{(2)}(Q^2)=\int\limits _{u_0^K}^1du~\left( \frac23 \varphi_K(u,\mu)+
\frac13 \varphi_K(\bar{u},\mu)\right)
\exp\left( -\frac{\bar{u}Q^2}{uM^2}+\frac{um_K^2}{M^2}\right)\,.
\label{LCSRKtw2}
\ee
The DA $\varphi_K(u,\mu)$ is defined as in Eq.~(\ref{varphi}),
with $q_1=u$ and $q_2=s$, so that $\bar{u}$ is the momentum 
fraction of the $s$ quark. In the above, the lower limit  
$u_0^K$ is related to the duality threshold
in the kaon channel $s_0^K$  by the equation 
$s_0^K=\bar{u}_0^K(Q^2/u_0^K+m_K^2)$ which should be solved
with $O(m_K^2)$ accuracy.
The $O(\alpha_s)$ correction to the twist 2 contribution 
has been calculated in Ref.~\cite{BKM} 
in the chiral limit. To obtain $F_K^{(2,\alpha_s)}$
we replace $\varphi_\pi$ by $\varphi_K$
in the expression for $F_\pi^{(2,\alpha_s)}$ 
given in App.~\ref{AppC}. In addition, there are $SU(3)_{fl}$ 
violating corrections in the hard amplitude. 
Note that the first-order in $m_s$ 
corrections are absent due to chirality. 
Nevertheless, indirectly,  $O(m_s)$  contributions 
will appear due to purely kinematical terms $O(p^2=m_K^2)$. 
To obtain these terms one has to recalculate the 
$O(\alpha_s)$ diagrams retaining $p^2 \neq 0$, which 
is beyond the task of this paper. 

The twist 3 and 4 terms in Eq.~(\ref{contribK})
can be cast in the same form
as the twist 4 contribution to the pion form factor:  
\ba
F_K^{(3,4)}(Q^2)= \int\limits ^1_{u_0^K}du 
\frac{\varphi^{(3,4)}_K(u,\mu)}{uM^2}
\exp\left( -\frac{\bar{u}Q^2}{uM^2} +\frac{u m_K^2}{M^2}\right) 
\nonumber
\\
+ \left(\frac1{Q^2+s_0^K}+\frac{(Q^2-s_0^K)m_K^2}{
(Q^2+s_0^K)^3} \right)\varphi^{(3,4)}_K (u_0^K,\mu)
e^{-s_0^K/M^2 +m_K^2/M^2}\,,
\label{tw4resK}
\ea
where  
\be
\varphi^{(3)}_K(u,\mu)= \frac{2 m_s f_{3K}}{3f_Ku} 
\int\limits^{u}{\cal D}\alpha_i \varphi_{3K}(\alpha_i)
\label{phiK3}
\ee
and 
\ba
\varphi^{(4)}_K(u,\mu)= 
-\frac{1}3\Bigg(4\left[2g_{1K}(u)+g_{1K}(\bar{u})\right]
-4\left[2G_{2K}(u)+G_{2K}(\bar{u})\right]-
2u\left[2g_{2K}(u)-g_{2K}(\bar{u})\right]
\nonumber
\\
+\int\limits^{u}{\cal D}\alpha_i
\Bigg[(1-2v)\left(2[2\varphi_{\perp K}(\alpha_i)-
\varphi_{\perp K}(\bar{\alpha_i})]
+\left[2\varphi_{\parallel K}(\alpha_i)-\varphi_{\parallel K}(\bar{\alpha_i})\right] \right)
\nonumber
\\
\!\!\!\!\!-2\left[2\widetilde{\varphi}_{\perp K}(\alpha_i)+
\widetilde{\varphi}_{\perp K}(\bar{\alpha}_i)\right]\!-\!
\left[2\widetilde{\varphi}_{\parallel K}(\alpha_i)
+\widetilde{\varphi}_{\parallel K}(\bar{\alpha}_i)\right]\!\Bigg]\Bigg).
\ea
In the surface term in Eq.~(\ref{tw4resK})
only $O(m_K^2)$ terms are
taken into account in accordance with our approximation.
Since the twist 3 contribution to the correlation function 
is proportional to $m_s$ it is  consistent to use 
the $SU(3)_{fl}$ limit of the twist 3 DA, in particular 
$\varphi_{3K}=\varphi_{3\pi}$ in Eq.~(\ref{phiK3}).  
Finally, for simplicity we adopt $SU(3)_{fl}$ limit for the 
numerically small twist 6 factorizable term.

\begin{figure}[t]
\begin{center}
\includegraphics[width=0.7\textwidth,angle=0]{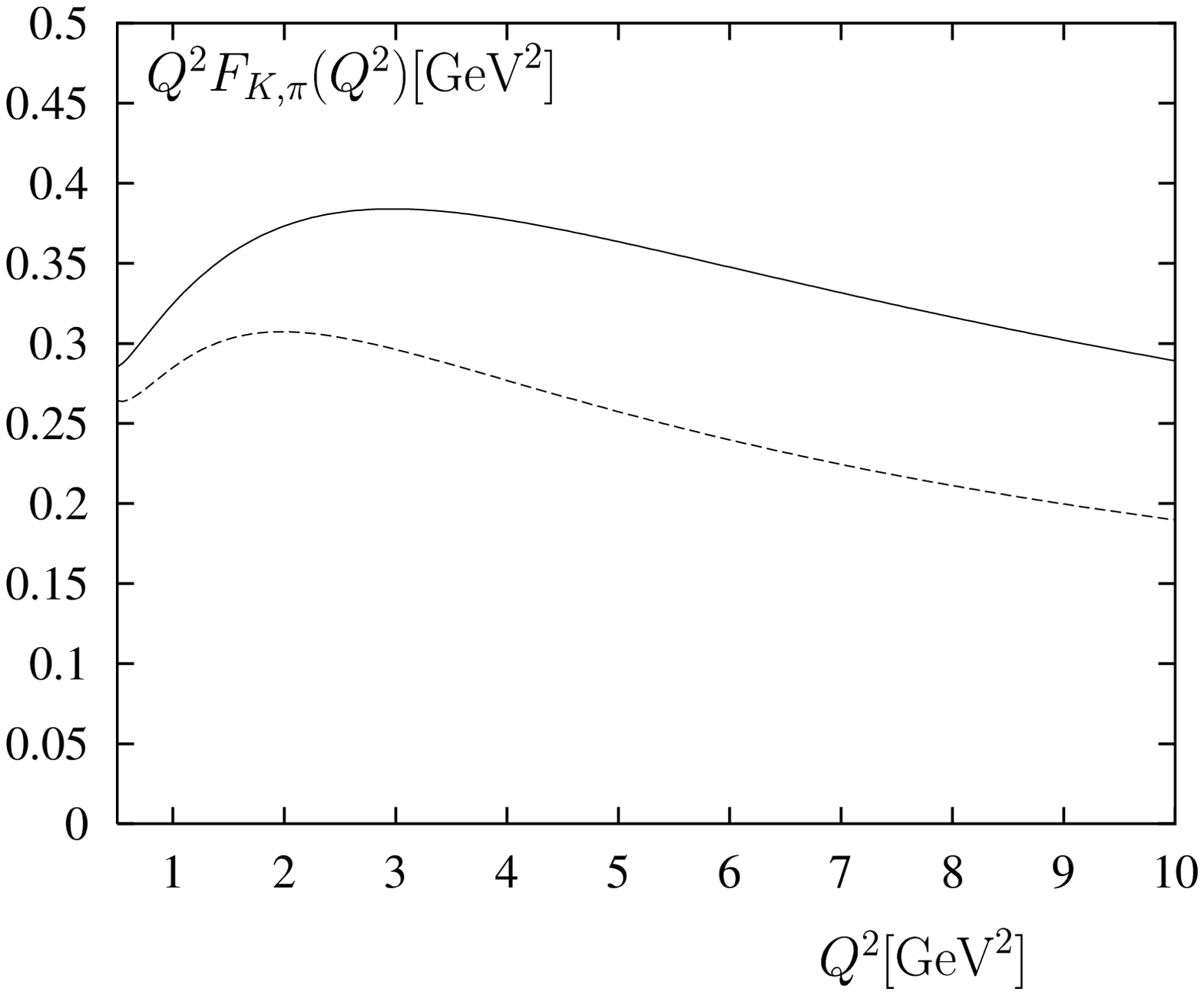}
\end{center}
\caption{ {\it  
The  LCSR prediction for the 
charged kaon  e.m. form factor $F_K(Q^2)$ 
(solid line) in comparison with the pion 
form factor obtained
with the asymptotic pion DA (dashed line) 
at $M^2= 1$ GeV $^2$. }
\label{fig5}}
\end{figure}

In Fig.~\ref{fig5}  we plot the kaon e.m. form factor calculated
with the following choice of parameters:

a) the twist 2 DA 
is taken with $a_1^K(\!\mbox{1 GeV})=-0.17$ (as estimated 
from the 2-point sum rule in Ref.~\cite{CZ84}) and neglecting all 
higher Gegenbauer coefficients (in particular $a_2^K=0$), 
that is, maximally close to the asymptotic regime; 

b) $s_0^K=1.2$ GeV$^2$ is determined from the two-point 
QCD sum rules for $f_K$ \cite{Pivovarov}. Importantly,
$s_0^K$ is larger than $s_0^\pi$ reflecting
heavier states in the kaon channel; 

c) for the strange quark mass we adopt  an interval $m_s(1 \mbox{GeV})=150 \pm 50$ 
MeV; 

d) the parameters of twist 3,4 kaon DA taken in 
the $SU(3)_{fl}$ limit\footnote{ As noted above 
this is only consistent for the 
twist 3 part of the sum rule. The accuracy of the 
twist 4 part can be improved further if one determines the 
nonperturbative parameters entering the kaon twist 3,4 DA
from the corresponding two-point sum rules in the kaon channel 
taking into account $SU(3)_{fl}$ violation, a task for future work.}:
$f_{3K}(1 \mbox{GeV})=f_{3\pi}(1 \mbox{GeV})=0.0035$ GeV$^2$, 
$\delta^2_K(1 \mbox{GeV})=\delta^2_\pi(1 \mbox{GeV})=0.17 \pm 0.05$
GeV$^2$ (see Eq.~(\ref{delta})), $\omega_{3K}(1 \mbox{GeV})= 
\omega_{3\pi}(1 \mbox{GeV})=-2.88$,
$\epsilon_{4K}(1 \mbox{GeV})=\epsilon_{\pi}(1 \mbox{GeV})=0.5$ 
\cite{BF,Ball}. The typical accuracy of all parameters except
$\delta_\pi^2$ is about 50\%. We also use $f_K = 1.22 f_\pi$.

Comparing the LCSR prediction for the  pion and kaon form factors 
calculated at $M^2=1$ GeV$^2$ 
we observe a noticeable $SU(3)_{fl}$ violating difference.
The ratio $F_K(Q^2)/F_\pi(Q^2)$ approaches 1.5 at $Q^2\sim10$ GeV$^2$.
A closer look at Eq.~(\ref{LCSRKtw2})
reveals that this difference 
originates from an interplay of two opposite effects. The  
$SU(3)_{fl}$ asymmetry due to $a_1\neq 0 $ in $\varphi_K(u)$ 
tends to suppress the kaon 
form factor because in the larger contribution (corresponding to the 
u-quark interacting with the virtual photon)
$\varphi_K(u)<\varphi_\pi(u)$ in the end-point integration region $u_0^K< u< 1$. 
On the other hand, the fact that the duality threshold for the kaon is higher, 
$s_{0K}>s_{0\pi}$, implies that 
the end-point region for the kaon form factor is itself larger, thereby
increasing $F_K$. Numerically, the latter effect 
turns out to be more important. Interestingly, the twist 3 contribution 
which is entirely an $SU(3)_{fl}$ violating effect is 
negligibly small, so that the $m_s$ uncertainty is unimportant. 
Note that in the ratio of kaon and pion form factors   
some theoretical uncertainties (e.g., due to $M^2$- and
scale-dependence) cancel leaving 
the major uncertainty in the Gegenbauer coefficient $a_1$. The unaccounted $SU(3)_{fl}$ 
violating effects in higher twists can presumably be neglected
within the present accuracy. If the kaon e.m. form factor is measured
one would then be able to constrain/fit $a_1$. 

Our final comment in this section concerns
the neutral kaon e.m. form factor. It can be easily 
calculated from the same LCSR (\ref{contribK})
if one replaces $u$-quark by $d$-quark in the initial correlation
function, having in mind that DA of $K^0$ and $K^+$ are equal due to
isospin symmetry. In particular, the leading twist 2 
contribution is obtained replacing the u-quark charge 
2/3 in Eq.~(\ref{LCSRKtw2}) by the d-quark charge -1/3. As a result
$F_{K^0}(Q^2)$ is a pure $SU(3)_{fl}$ violating effect proportional to the integral 
over the difference $\varphi_K(u)-\varphi_K(\bar{u})$. 
The numerical result is small : 
$Q^2F_{K^0}(Q^2)= 0.05-0.09$ GeV$^2$, (at $1< Q^2<10$ GeV$^2$) implying that 
the measurement of this form factor is a difficult task. 
 
We conclude that the LCSR method allows to systematically account  for
$SU(3)_{fl}$ breaking effects in the kaon form factors, and that 
these effects revealed by the ratio 
of $K^+$ and $\pi^+$ form factors are predicted to be quite noticeable.

\section{The $K\to \pi$ form factor}  
\label{Sect6}

As a final application of LCSR in this paper we consider 
the $K\to \pi$ form factor
$f^+_{K\pi}$ defined as 
\be
\langle \pi^-(p-q) \mid \bar{s}\gamma_\mu 
u \mid K^0(p)\rangle =2f^+_{K\pi}(q^2) p_\mu -
(f^+_{K\pi}(q^2) - f^-_{K\pi}(q^2))q_\mu  \,.
\ee
As explained in Sect.~2 this form factor can  
be calculated from the correlation 
function (\ref{2:cor}) in two different ways, either 
from the vacuum-to-pion or from the vacuum-to-kaon 
correlation functions.
Both calculations are  valid only at 
sufficiently large  spacelike momentum
transfer $Q^2\geq $ 1 GeV$^2$, whereas 
the  form factor $f^+_{K\pi} $ is measurable only 
at timelike momenta where the LCSR method is not applicable, 
e.g., at $0 < q^2 < (m_K-m_\pi)^2$ 
in the $K_{e3}$ decays, or at 
$(m_K+m_\pi)^2 < q^2 <m_\tau^2$ in $\tau \to K\pi \nu$ 
decays.

Nevertheless, one is able to use the fact that 
LCSR obtained for two different settings yield 
one and the same physical parameter and derive 
useful constraints on the light-cone DA of the pion 
and kaon involved in both sum rules. To explain
the idea we explicitly write down the LCSR
obtained from the vacuum-to-pion correlator:
\be
f_{K\pi}^+(Q^2)=\frac{f_\pi}{f_K}\int_{u_0^K}^1
du~\varphi_\pi(u,\mu)\exp\left( 
-\frac{\bar{u}Q^2}{uM^2}+\frac{m_K^2}{M^2}\right) + ...\,.
\label{LCSRKpi1}
\ee
Using instead the vacuum-to-kaon correlator, one gets 
\be
f_{K\pi}^+(Q^2)=\frac{f_K}{f_\pi}\int_{u_0^\pi}^1
du~\varphi_K(\bar{u},\mu)
\exp\left( -\frac{\bar{u}Q^2}{uM^2}-\frac{\bar{u}m_K^2}{M^2}\right)+...
\,,
\label{LCSRKpi2}
\ee
In the region $1<Q^2 < 3$  GeV$^2$ and at $M^2\sim 1 $ GeV$^2$ 
both sum rules are valid and the higher twist 
and $O(\alpha_s)$ contributions denoted by ellipses
are small, so that we may neglect them for the sake of simplicity.
Equating (\ref{LCSRKpi1}) and (\ref{LCSRKpi2})
in this region one may constrain the nonasymptotic
coefficients. In particular, the rate of $SU(3)_{fl}$
breaking  asymmetry in the kaon DA $\varphi_K$ can be estimated if 
the pion DA $\varphi_\pi$ is determined with a sufficient accuracy.

\begin{figure}[t]
\begin{center}
\includegraphics[width=0.7\textwidth,angle=0]{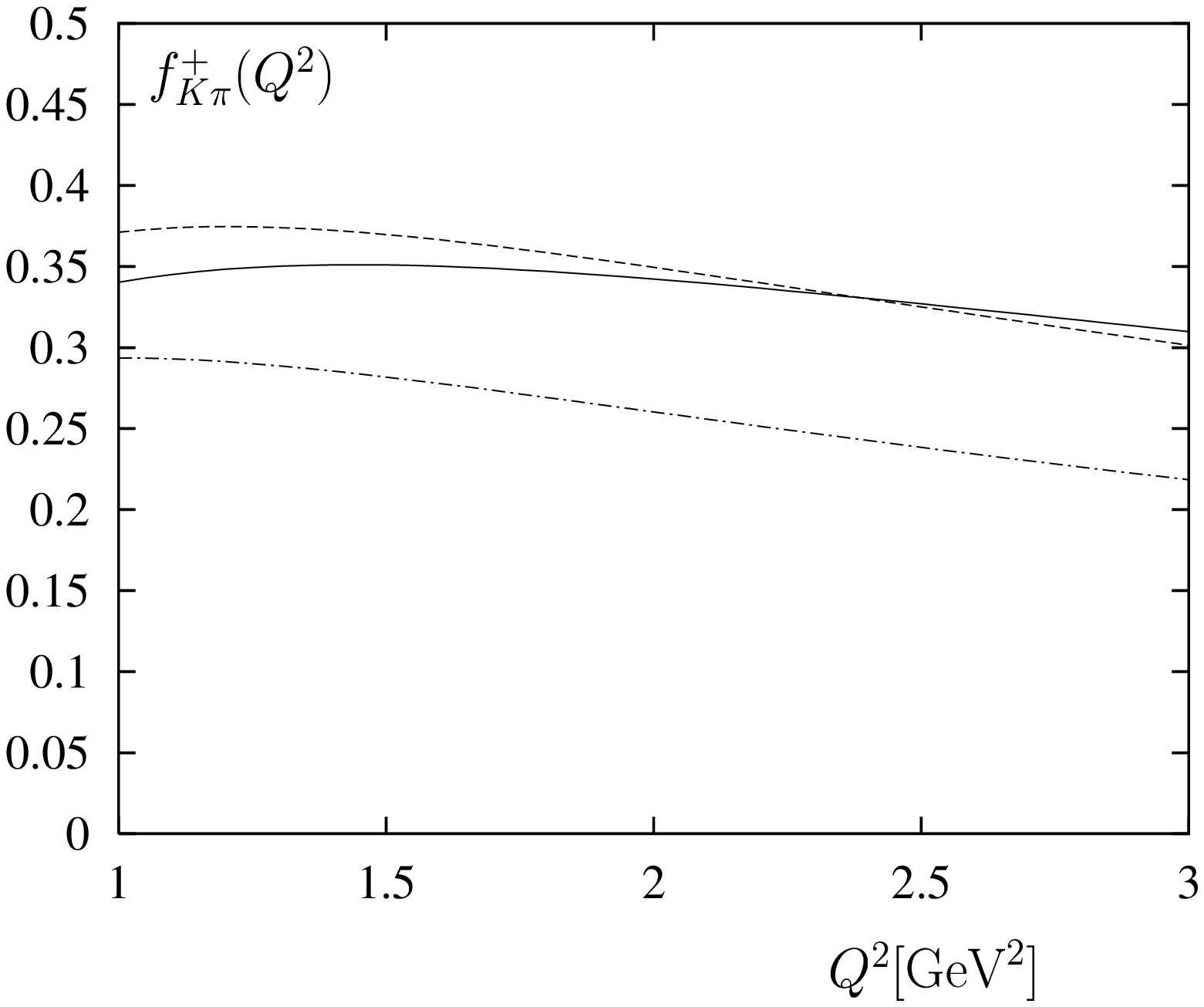}
\end{center}
\caption{ {\it  
The $K\to \pi$ form factor at spacelike region 
calculated with twist 2 accuracy:
from Eq.~(\ref{LCSRKpi1}) with the asymptotic
pion DA (solid) and  from Eq.~(\ref{LCSRKpi2})
with the kaon DA including the $SU(3)_{fl}$ 
violating correction  $\sim a_1$ (dashed) and 
at $a_1=0$ (dash-dotted).}
\label{fig6}}
\end{figure}

As a numerical illustration, in Fig.\ref{fig6}
we compare the r.h.s. of Eqs.~(\ref{LCSRKpi1})
and (\ref{LCSRKpi2}) calculated respectively 
with the asymptotic pion DA 
and with the kaon DA adopted in our calculation of $F_K$
in the previous section,
that is with $a_1(1 \mbox{GeV})=-0.17$ and all $a_{n>1}=0$.
The resulting agreement of two different sum rules 
is nontrivial and ensures confidence in the whole procedure 
and especially in the choice of duality thresholds in both pion and 
kaon channels. Note that the agreement is violated if we put $a_1=0$.

\section{Conclusions} 
\label{Conclusions}

In this paper, we have studied 
the correlation functions of light-quark currents used 
to derive the LCSR for the pion and kaon form factors.
We have demonstrated that the Ward identities
for these correlators yield relations between DA of twist 4, 
a new alternative to using the QCD equation of motions.
On the phenomenological side, we have corrected
the expression for the twist 4 contribution to the LCSR for the pion 
form factor. The form factor calculated 
with the purely asymptotic pion DA is generally consistent 
with the recent Jefferson Lab data. On the other hand, 
constraining the nonasymptotic part of the pion twist 2 DA in terms
of separate Gegenbauer coefficients 
demands more data at intermediate momentum transfers, $1 <Q^2 <10$ GeV$^2$
and largely depends on the particular ansatz adopted for this DA. 
A recent study of a similar problem for the 
$\gamma^* \gamma ^* \to \pi^0$ form factor  can be found in 
Ref.~\cite{Diehletal}.

We have presented the first LCSR prediction
for the kaon e.m. form factor and demonstrated
that within the sum rule approach the $SU(3)_{fl}$ violating difference between kaon
and pion form factors is systematically calculable in powers of the strange quark mass. 
It has been shown that a useful complementary information
concerning the kaon DA can be obtained  
from the comparison of two independent sum rules for the 
$K \to \pi$ form factor. In general, our results support the point of view that 
LCSR for the pion and kaon form factors combined with sufficiently precise 
data on these form factors represent a very useful tool 
for probing the pion and kaon light-cone
distribution amplitudes.

\section*{Acknowledgments}

The authors thank Patricia Ball, Vladimir Braun and Johann K\"uhn 
for useful 
discussions and comments. The work of A.K. is supported 
by BMBF (Bundesministerium
f\"ur Bildung und Forsching). He also
acknowledges the support of STINT (Swedish Foundation
for International Cooperation in Research and Higher Education) 
during his visit to the Lund University where this project was 
initiated. The work of JB is supported by the Swedish Research Council
and by TMR, EC--Contract No.
ERBFMRX--CT980169 (EURODAPHNE)

\appendix
\renewcommand{\theequation}{\Alph{section}\arabic{equation}}
\section{Light-cone expansion of the quark propagator}
\setcounter{equation}{0}
\label{AppA}

The expansion of the quark propagator with a nonzero mass 
$m$ near the light-cone $(x_1-x_2)^2=0$ reads 
(see e.g. Ref.~\cite{BBKR}): 
\ba
S(x_1,x_2,m)&&\equiv -i\langle 0 | T \{q(x_1)\bar{q}(x_2)\}| 0 \rangle
\nonumber
\\
&&=\int\frac{d^4k}{(2\pi)^4}e^{-ik(x_1-x_2)}\Bigg\{
\frac{\not\!k +m}{k^2-m^2}
-\int\limits_0^1 dv\, G^{\mu\nu}(vx_1+(1-v)x_2)
\nonumber
\\
&&
\times \Big[ \frac12 \frac {\not\!k +m}{(k^2-m^2)^2}\sigma_{\mu\nu} -
\frac1{k^2-m^2}v(x_1-x_2)_\mu \gamma_\nu \Big]\Bigg\}\,, 
\ea
with $G^{\mu\nu}= g_sG_{\mu\nu}^a(\lambda^a/2)$, 
$\mbox{Tr}(\lambda^a\lambda^b)=2\delta^{ab}$. 
At $m=0$, after  the integration over $k$ this expression
reduces to the propagator 
derived in Ref.~\cite{BB8889}, (see also Ref.~\cite{BKM}): 
\ba
S(x_1,x_2,0)
=\frac{\not\!x_1-\not\!x_2}{2\pi^2[(x_1-x_2)^{2}]^2}  
-\frac{1}{16\pi^2(x_1-x_2)^2}
\int\limits_0^1 dv\, G^{\mu\nu}(vx_1+(1-v)x_2)
\nonumber
\\
\times \Big[(\not\!x_1-\not\!x_2)\sigma_{\mu \nu} -4iv(x_1-x_2)_\mu
\gamma_\nu\Big]\,.
\ea


\section{Light-cone distribution amplitudes}
\setcounter{equation}{0}
\label{AppB}

The light-cone DA 
of the pseudoscalar meson $P=\pi,K$ are defined according to 
Refs.~\cite{BF,Ball}. The 
matrix element of the axial-vector bilocal operator  
is expanded around the light-cone ($x_1^2=x_2^2=(x_1-x_2)^2=0$): 
\begin{eqnarray}
\langle 0| \bar{q}_2(x_2) \gamma_{\mu} \gamma_5 q_1(x_1) | P(p) \rangle 
&=& f_P \int\limits_0^1 du e^{-iupx_1-i\bar{u}px_2}\Big\{
i p_{\mu} \left(\varphi_{P}(u) + (x_1-x_2)^2 g_{1P}(u) \right)
\nonumber \\
&& +\left((x_{1} -x_{2})_\mu- 
\frac{p_{\mu}(x_1-x_2)^2}{p(x_1-x_2)}\right)g_{2P}(u)\Big\} \,,
\label{varphi}
\end{eqnarray}
retaining the leading twist 2 DA  $\varphi_P(u)$ and 
the twist~4 DA $g_{1P}(u)$ and $g_{2P}(u)$,
where  $u$ is the light-cone momentum fraction of the quark $q_1$.
From the local limit  of Eq.~(\ref{varphi})
one has the following normalization conditions:
\be
\int_0^1 du \,\varphi_P(u) =1\,,~~~~
\int_0^1 du \,g_{2P}(u)=0\,.
\ee
The twist~3 quark-antiquark DA $\varphi_{p P}$ and 
$\varphi_{\sigma P}$
and the quark-antiquark-gluon DA $\varphi_{3P}$ are defined as follows:
\begin{eqnarray}
\langle 0|\bar{q}_2(x_2) i\gamma_5 q_1(x_1)|P(p)\rangle\!\! &=&\!\! 
f_P \mu_P\!\int\limits_0^1 \!du\, e^{-iupx_1-i\bar{u}px_2} \varphi_{p
  P}(u)\;,
\nonumber \\
\langle 0|\bar{q}_2(x_2) \sigma_{\alpha\beta}\gamma_5 q_1(x_1)|P(p)\rangle 
\!\!&=&\!\! \frac{i f_P \mu_P}{6}\left(
1-\frac{m_P^2}{\mu_P^2}\right)
\Big[p_\alpha (x_1-x_2)_\beta\! -\! 
p_\beta (x_1 - x_2)_\alpha\Big]
\nonumber \\
\times\int\limits_0^1\! du\, e^{-iupx_1-i\bar{u}px_2} 
\varphi_{\sigma P}(u)\;,
\end{eqnarray}
where $\mu_P = m_P^2/(m_{q_1}+m_{q_2})$, and 
\ba 
&&\langle 0 |\bar{q}_2(x_2)
\sigma_{\mu\nu}\gamma_5 G_{\alpha\beta}(vx_1 +\bar{v}x_2) q_1(x_1)
| P(p) \rangle 
=if_{3P}\Big[(p_\alpha p_\mu g_{\beta\nu}-p_\beta p_\mu g_{\alpha\nu})
\nonumber\\
&&-(p_\alpha p_\nu g_{\beta\mu}-p_\beta p_\nu g_{\alpha\mu})\Big]
\int{\cal D}\alpha_i\,\varphi_{3P}(\alpha_i,\mu)
e^{-i\alpha_1px_1-i\alpha_2 px_2 -i\alpha_3(vpx_1+\bar{v}px_2)}\,,
\label{tw3matr}
\ea
all these DA being normalized to unity.
Furthermore, the  quark-antiquark-gluon twist~4 DA are defined by 
the following matrix elements:
\begin{eqnarray}
&&\langle 0| \bar{q}_2(x_2)\gamma_\mu \gamma_5 G_{\alpha\beta}(vx_1+\bar{v}x_2)
q_1(x_1) |P(p)\rangle = f_P \int\! {\cal D}\alpha_i  
e^{-i\alpha_1 px_1-i\alpha_2px_2 -i\alpha_3(vpx_1+\bar{v}px_2)}
\nonumber
\\
&&\times \Big\{ p_\mu \frac{p_\alpha (x_1-x_2)_\beta-p_\beta
(x_1-x_2)_\alpha}{p(x_1-x_2)}\varphi_{\parallel P}(\alpha_i) 
+ ( g^\perp_{\mu\alpha}p_\beta  - g^\perp_{\mu\beta}p_\alpha)
\varphi_{\perp P}(\alpha_i)\Big\}\,, 
\label{tw431}
\end{eqnarray}
\begin{eqnarray}
&&\langle 0| \bar{q}_2(x_2)\gamma_\mu i\widetilde{G}_{\alpha\beta}(vx_1+\bar{v}x_2)
q_1(x_1) |P(p)\rangle = f_P \int\! {\cal D}\alpha_i  
e^{-i\alpha_1 px_1-i\alpha_2px_2 -i\alpha_3(vpx_1+\bar{v}px_2)}
\nonumber
\\
&&\times \Big\{ p_\mu \frac{p_\alpha (x_1-x_2)_\beta-p_\beta
(x_1-x_2)_\alpha}{p(x_1-x_2)}\widetilde{\varphi}_{\parallel P}(\alpha_i) 
+ ( g^\perp_{\mu\alpha}p_\beta  - g^\perp_{\mu\beta}p_\alpha)
\widetilde{\varphi}_{\perp P}(\alpha_i)\Big\}\,, 
\label{tw432}
\end{eqnarray}
where $\widetilde G_{\alpha\beta}= \frac 12 \epsilon_{\alpha\beta\rho\lambda}
G^{\rho\lambda}$ and the following  abbreviations are used: 
$${\cal D} \alpha_i = d\alpha_1d\alpha_2d\alpha_3
\delta \left(1\!-\! \alpha_1\! -\! \alpha_2 \!-\! \alpha_3 \right)
~~\mbox{and}~~ 
g_{\alpha\beta}^\perp = g_{\alpha\beta} - 
\frac{(x_1-x_2)_\alpha p_\beta + (x_1-x_2)_\beta p_\alpha}{p(x_1-x_2)} ~\,.
$$

The distribution amplitudes are constructed \cite{BF}
using the formalism of the conformal expansion. The most familiar example is  
the  twist 2 DA \cite{exclusive}
\begin{equation}
\varphi_P(u,\mu ) = 6 u \bar u \left[ 1 + 
\sum_{n=1}a_{n}^P(\mu) C_{n}^{3/2}(u - \bar u) \right],
\label{phipi}
\end{equation}
where the  expansion 
goes in Gegenbauer polynomials $~C_{n}^{3/2}$, 
the first four polynomials being
\begin{eqnarray}
C_1^{3/2}(x) &=& 3x\,,~~~
C_2^{3/2}(x) = -\frac{3}{2}(1-5 x^2)\,,
\nonumber \\
C_3^{3/2}(x) &=& -\frac{5}{2}x(3 -7x)\,,~~~
C_4^{3/2}(x) = \frac{15}{8}(1-14x^2+ 21x^4)\,.
\end{eqnarray}
The scale-dependence is given in the leading order  by
\begin{equation} 
a_n^{P}(\mu_2)=\left[L(\mu_2,\mu_1)\right]^{-\gamma_n^{(0)}/\beta_0}
a_n^{P}(\mu_1)~,
\label{anom}
\end{equation}
where $L(\mu_2,\mu_1)=\alpha_s(\mu_2)/\alpha_s(\mu_1)$, 
$\beta_0=11- \frac23 N_F$ and the anomalous dimensions are
\begin{equation}
\gamma_{n}^{(0)}=C_F\left[3 +\frac{2}{(n+1)(n+2)}-4\left(\sum_{k=1}^{n+1}
\frac1k\right)\right]\,.
\label{gamman}
\end{equation}
For the pion, 
the coefficients $a_n^\pi$ vanish at odd $n$ in the isospin 
symmetry limit.

The twist 3 and ~4 DA have been derived in Ref.~\cite{BF}
using QCD equations of motion and conformal expansion.
In Ref.~\cite{Ball} the meson mass correction have 
been worked out. We present here the explicit 
expression for these DA to next-to-leading 
accuracy in the conformal spin, including the meson mass 
corrections
(as explained in more detail in \cite{BF,Ball}) 
and using the original notations of Ref.~\cite{BF}.
Note that $SU(3)_{fl}$ violating nonasymptotic
corrections to these DA (analogous to $a_1\neq 0$ for $\varphi_K$) 
are still missing and have to be worked out in future.  

The twist 3 DA of the pseudoscalar meson, to next-to-leading order in conformal spin read:
\ba
\qquad \varphi_{p P}(u)\!\!\!&=& \!\!1+ 
\left(30\frac{f_{3P}}{\mu_P f_P} -\frac{5}2\frac{m_P^2}{\mu_P^2}
\right)C_2^{1/2}(2u-1)
\nonumber 
\\
\!\!\!&&\!\!+
\left(-3\frac{f_{3P}\omega_{3P}}{\mu_P f_P} -\frac{27}{20}\frac{m_P^2}{\mu_P^2}(1+6a_2^P)
\right)C_4^{1/2}(2u-1)\,,
\\
\qquad \varphi_{\sigma P}(u)\!\!\! &=& \!\!6u\bar u 
\Bigg\{1\!+\!\Big(5\frac{f_{3P}}{\mu_P f_P}(1\!-\!\frac1{10}\omega_{3P}) 
-\frac{7}{20}\frac{m_P^2}{\mu_P^2}(1\!+\!\frac{12}{7}a_2^P)\Big)
C_2^{3/2}(2u\!-\!1)\!\Bigg\},
\\
\varphi_{3P}(\alpha_i)\!\!\!&=&\!\! 360\alpha_1 \alpha_2\alpha_3^2\left(1+
\frac{\omega_{3P}}{2}(7\alpha_3-3)\right)\,.
\ea
The nonperturbative parameter $f_{3P}$ is given by 
the matrix element which corresponds to the local limit in  Eq.~(\ref{tw3matr}).
The second parameter $\omega_{3P}$ determining the nonasymptotic 
parts of twist 3 DA is defined with the following matrix element
(up to higher twist corrections):
\be
\langle 0 |\bar{q}_2\sigma_{\mu\lambda}\gamma_5
[D_\beta,G_{\alpha\lambda}] q_1-
\frac37 \partial_\beta\bar{q}_2\sigma_{\mu\lambda}\gamma_5
G_{\alpha\lambda} q_1
|P(p)\rangle= \frac{3}{14}f_{3P}\omega_{3P}p_\alpha p_\beta p_\mu\,. 
\ee
The scale dependence of the twist 3 parameters is given by:
\begin{eqnarray}
\mu_{P}(\mu_2) = \left[L(\mu_2,\mu_1)\right]
^{-\frac{4}{\beta_0}}\mu_{P}(\mu_1)\,,~~~
f_{3P}(\mu_2) = \left[L(\mu_2,\mu_1)\right]^
{\frac{1}{\beta_0}\left(\frac{7 C_F}{3}+N_c\right)}
f_{3P}(\mu_1)\,,
\\
(f_{3P}\omega_{3P})(\mu_2) = 
\left[L(\mu_2,\mu_1)\right]^{\frac{1}{\beta_0}\left(\frac{7 C_F}{6}+\frac{10N_c}{3}\right)}
(f_{3P}\omega_{3P})(\mu_1)\,.
\end{eqnarray}
Finally, the four twist 4 three-particle DA defined in
Eq.~(\ref{tw431}), (\ref{tw432}) are:
\begin{eqnarray}
\!\!\!\varphi_{\parallel P}(\alpha_i)\!\! &=& \!\!\!120(\delta_P^2 \epsilon_P-
\frac{9}{20}a_{2}^Pm_P^2)(\alpha_1-\alpha_2)\alpha_1\alpha_2\alpha_3 \,,
\nonumber \\
\!\!\!\varphi_{\perp P}(\alpha_i)\!\! &=&\!\!\! 30 (\alpha_1-\alpha_2)
\alpha_3^2 \left[\frac{\delta^2_P }{3} +2\delta^2_P \epsilon_P
(1-2\alpha_3)+\frac{9}{40}a_{2}^Pm_P^2(1-\alpha_3)\right] ,
\nonumber \\
\!\!\!\widetilde \varphi_{\parallel P} (\alpha_i) 
\!\!&=&\!\! - 120 \delta_P^2 \alpha_1\alpha_2 \alpha_3
\left[ \frac{1}{3} + \epsilon_P(1-3\alpha_3) \right],
\nonumber \\
\!\!\!\widetilde \varphi_\perp(\alpha_i)\!\! &=&\! \!\!30 \alpha_3^2
\left[\left(\frac{\delta_P^2 }{3} + 
2 \delta_P^2 \epsilon_P(1-2\alpha_3)\right)\!(1\!-\!\alpha_3)
\!+\!\frac{9}{40}a_2^Pm_P^2
(\alpha_1^2+\alpha_2^2-4\alpha_1\alpha_2)\right]\!,
\label{tw43p}
\end{eqnarray}
normalized as 
\be
\int\! {\cal D}\alpha_i \varphi_{\parallel P}(\alpha_i)=
\int\! {\cal D}\alpha_i \varphi_{\perp P}(\alpha_i)=0\,,
~~-\!\!\int\! {\cal D}\alpha_i \widetilde\varphi_{\parallel P}(\alpha_i)=
\int\! {\cal D}\alpha_i \widetilde\varphi_{\perp P}(\alpha_i)=
\frac{\delta_P^2}{3}\,.
\ee
The corresponding two-particle DA have the following expressions:
\begin{eqnarray}
g_{1P}(u) =\frac{5}2\delta_P^2u^2\bar{u}^2+
\Bigg\{ \frac{f_{3P}
m_P^2}{4f_P\mu_P }
\Big[30(1-2u\bar{u})-\omega_{3P}\left(3-u\bar{u}(27-56u\bar{u})\right)\Big]
\nonumber
\\
+\frac{m_P^2}{320}\Big[5(25-29u\bar{u})-12a_2^P\Big(1-5u\bar{u}
(19-52u\bar{u})\Big)\Big]\Bigg\}u\bar{u}
\nonumber
\\
+\frac{1}{2}(\delta_P^2\epsilon\!-\!\frac{9}{20}a_2^Pm_P^2)
\Big[ 2u^3(10\!-\!15u\!+\!6u^2)\ln u\! +\!
2\bar{u}^3(10\!-\!15\bar{u}\!+\!6\bar{u}^2)\ln\bar{u}
\!+\!u\bar{u}(2\!+\!13u\bar{u})\Big]
\label{tw4g1}
\end{eqnarray}
\begin{eqnarray}
g_{2P}(u)= \Bigg[
\frac{10}{3}\delta^2_P+
m_P^2\left(1+\frac98a_2^P(1\!-\!7u\bar{u})\right)\!-\!
\frac{f_{3P}m_P^2}{f_P\mu_P }\left(10\!-\!\omega_{3P}(1\!-\!7u\bar{u})\right)
\Bigg]\bar u u(u\!-\!\bar u)\,. 
\label{tw4g2}
\end{eqnarray}
The nonperturbative parameter $\delta^2_P$ is defined as 
\begin{equation}
\langle 0 |\bar{q}_2\widetilde{G}_{\alpha\mu}\gamma^\alpha q_1
|P(p) \rangle=-i\delta^2_P f_P p_\mu \,,
\label{delta1}
\end{equation}
with the scale-dependence:
\begin{eqnarray}
\delta^2_P(\mu_2) = \left[L(\mu_2,\mu_1)\right]
^{\frac{8C_F}{3\beta_0}}\delta^2_P(\mu_1)\,,
\end{eqnarray}
whereas the second parameter $\epsilon_P$  
determining the nonasymptotic corrections has the 
following definition in terms of a local matrix element 
(up to twist 5 corrections)\cite{BF,Ball}:
\be
\langle 0 |\bar{q}_2[D_\mu,\widetilde{G}_{\nu\xi}]\gamma^\xi q_1-
\frac49\partial_\mu\bar{q}_2\widetilde{G}_{\nu\xi}\gamma^\xi q_1
|P(p)\rangle= -\frac{8}{21}f_P\delta_P^2\epsilon_P\left(p_\mu 
p_\nu-\frac14m_P^2g_{\mu\nu}\right)\,.
\ee
The corresponding scale dependence is :
\be
(\delta_P^2\epsilon_P)(\mu_2) = 
\left[L(\mu_2,\mu_1)\right]
^{\frac{10N_c}{3\beta_0}}(\delta^2_P\epsilon_P)(\mu_1)\,.
\ee

\section{Radiative Corrections to the Twist 2 Pion Form Factor}
\setcounter{equation}{0}
\label{AppC}

Here, for completeness, we present the formula for the radiative correction
to the twist 2 part of the sum rule for the pion form factor
obtained in Ref.~\cite{BKM}:  

\begin{equation}
 F^{(2,~\alpha_s)}_{\pi}(Q^2) = \int\limits_{0}^1 \!du\, \varphi_{\pi}(u, \mu) 
\Bigg[ \Theta(u-u_0){ \cal F}_{\rm soft}(u,M^2,s_0)+
       \Theta(u_0-u){ \cal F}_{\rm hard}(u,M^2,s_0)\Bigg],
\label{3:NLO}
\end{equation}
where
\begin{eqnarray}
\lefteqn{
{\cal F}_{\rm soft}(u,M^2,s_0)=
}
\nonumber\\ 
&=&\frac{\alpha_s}{4\pi}C_F \Bigg \{
\exp \left( - \frac{\bar uQ^2}{u M^2} \right)
\Bigg[ 
\!-9 + \frac{\pi^2}{3} + 3\ln \frac{Q^2}{\mu^2} 
+ 3\ln\frac{\bar u Q^2}{u \mu^2}
-\ln^2 \frac{Q^2}{\mu^2} - \ln^2 \frac{\bar u Q^2}{u \mu^2}\Bigg]
\nonumber \\
&& +
\int\limits_{\bar u Q^2/u}^{s_0} \frac{ds\, Q^2 e^{-s/M^2}}{u (Q^2+s)^3}
\Bigg[ 5s +Q^2\left(1+2\ln \frac{-\rho}{\mu^2}\right)
+2\left(\frac{Q^2}{\bar u}+ s \right)\ln \frac{-\rho}{s}
\nonumber \\
&& + \frac{2Q^2}{u}\left(\frac{Q^2+s}{s} +\frac{2M^2+Q^2+s}{M^2}
\ln\frac{-\rho}{s} \right)\ln\frac{-\rho}{\mu^2} \Bigg]
\nonumber
 \\
&& +  \int\limits_0^{\bar u Q^2/u} \frac{ds\, Q^2 e^{-s/M^2}}{u \bar u (Q^2+s)^3}
\Bigg[ 2u \left(Q^2-s+s\ln\frac{s}{\mu^2}\right)
+\Bigg(-Q^2 + 5s+2(Q^2-s)\ln\frac{s}{\mu^2} 
\nonumber
\\
&&
\!-\frac{s(Q^2+s)}{M^2} \left(-3+2 \ln\frac{s}{\mu^2} \right)\Bigg)
\ln\frac{\rho}{\mu^2}\Bigg] 
+ 2\frac{u_0^2}{u^2} e^{-s_0/M^2}
\ln\frac{-\rho_0}{\mu^2}\ln \frac{u-u_0}{\bar u_0}\Bigg\}\,,
\label{3:soft2}
\end{eqnarray}
and
\begin{eqnarray}
\lefteqn{{\cal F}_{\rm hard}(u,M^2,s_0) =}
\nonumber\\
&=& \frac{\alpha_s}{4\pi}C_F \Bigg\{
\int\limits_0^{s_0}\frac{ds\, Q^2e^{-s/M^2}}{\bar u(Q^2+s)^3} 
\Bigg[ 2\left(Q^2 -s +s \ln\frac{s}{\mu^2}\right)
+\frac{1}{u }\Bigg(
-Q^2 + 5s+2(Q^2-s)\ln\frac{s}{\mu^2}
\nonumber
\\
&&-\frac{s (Q^2+s)}{M^2}
\left(-3+2 \ln\frac{s}{\mu^2}\right) \Bigg)\ln\frac{\rho}{\mu^2}\Bigg] 
- \frac{u_0\bar u_0}{u \bar u}e^{-s_0/M^2}
\left(2\ln\frac{s_0}{\mu^2}-3\right)
\ln\frac{\rho_0}{\mu^2} \Bigg\}.   
\label{3:hard2}
\end{eqnarray}
Here $\rho = \bar u Q^2 - us$ and $\rho_0 = \bar u Q^2 - us_0= (1-u/u_0)Q^2$.

\end{document}